\documentclass{du-journals}
\usepackage{amsthm,amsmath}
\usepackage{color}
\usepackage{subfigure}
\usepackage[noadjust,nospace]{cite}
\journal{Iranian Journal of Astronomy and Astrophysics}
\title{Bianchi I Cosmology and Scalar Vector Tensor Brans Dicke Gravity}
\subtitle{Bianchi I Cosmology and Scalar Vector Tensor Brans Dicke Gravity}
\author[1]{Hossein Ghaffarnejad$^*$}
\address[1]{Faculty of Physics, Semnan University, P.C. 35131-19111, Semnan, Iran; \\ $^*$email: hghafarnejad@semnan.ac.ir}

\author[2]{Hoda Gholipour}
\address[2]{Faculty of Physics, Semnan University, P.C. 35131-19111, Semnan, Iran; \\ email: gholipour.hoda@semnan.ac.ir}
\begin{document}
\begin{abstract}
We consider Brans Dicke scalar vector tensor gravity to study an
inflationary scenario of the accelerating expansion of the
universe for which the anisotropy property occurs. We study both
primordial inflation and late inflationary period of the universe.
To do so we, use Bianchi I line element where spatial part has
cylindrical symmetry along $x$ direction in the local Cartesian
coordinates. To seek stabilization of our obtained metric solution,
we apply dynamical system approach to obtain critical manifolds in
phase space and determine which of them confirms the inflation
with stable nature in presence of the anisotropy property of
spacetime. We solve dynamical equations for different directions
of the timelike dynamical vector field. We obtain several critical
manifolds whose nature of stable (sink) or quasi stable (saddle)
are dependent on the direction of the used vector field. At last, we
should point that observational constraint on the Brans Dicke
parameter $\omega>40000$ which satisfied by the well known Brans
Dicke scalar tensor gravity is not valid for our modified scalar
vector Brans Dicke gravity because of presence of timelike
dynamical vector field.
\end{abstract}

\begin{keywords}
Anisotropic cosmology, Bianchi model, Dark energy, Timelike vector
fields, Perfect fluids, Inflation.
\end{keywords}

\section{Introduction}
The $\Lambda CDM$ model which is confirmed by the standard
cosmology has a great success in explaining the observations of
the cosmic microwave background radiation (CMBR) temperature
anisotropy, as well as the galaxies distribution and whose motion
\cite{1,2,3,4}. This model which is based on the validity of the
cosmological principle (the spatial homogeneity and spatial
isotropy) and the Einstein`s general theory of relativity explain
most large-scale observations with unprecedented accuracy.
However, several directional anomalies have been reported in
various large-scale observations. In short these anomalies are
called as follows: the polarization distribution of the quasars
\cite{5}, the velocity flow \cite{6,7,8}, the handedness of the
spiral galaxies \cite{9,10,11}, the anisotropy of the cosmic
acceleration \cite{12,13,14,15,16,17}, the anisotropic evolution
of fine-structure constant \cite{18,19,20} and asymmetry of the
CMBR parity \cite{21,22,23,24,25}. In fact, origin of these
anomalies do not still understood and so they treat as puzzles.
There are two different proposals to understand these problems as
follows: The first perhaps is they are originated from cosmological
effects which should be described via alternative gravity theories
instead of the Einstein`s general theory of relativity. Other
possibility which arises these directional anomalies can be
systematic errors or contaminations of measuring instruments and
etc., which should be excluded from the future data analysis. In
the latter case, one usually accept validity of the standard
cosmological $\Lambda CDM$ model while in the former proposal one
use an alternative gravity model instead of the Einstein`s general
theory of relativity. Zhao and Santos, did provided full review
about these proposals \cite{26} where the directional anomalies
predict a preferred axis called as \textit{`Axis~of Evil`} in
large scale of the Universe. In short, they compared the preferred
directions in large-scale observations and the CMBR kinematic
dipole and found a strong alignment between them. In fact, CMB
radiation dipole is caused by motion of the solar system in the
universe which is a non-cosmological effect. For review on data
results of WMAP (Wilkinson-microwave anisotropy probe) and the
Planck satellite, one can see references
\cite{27,28,29,30,31,32,33,34,35,36,37,38,39}.
 However, some alternative cosmological models are provided to satisfy these anomalies where
the cosmological principle (spatial homogeneity and spatial
isotropy in large scales structure) should be violated. In
general, anisotropic curved spacetimes should be supported by
anisotropic stress energy tensor of matter fields which are not
present in the standard FLRW cosmology \cite{40,41}.  For
anisotropic vector field models, one can see \cite{42,43} and for
anisotropic cosmological constant with dark energy
\cite{44,45,46,47,48,49,50,51,52,53,54,55,56,57}. See also
\cite{53,54,55} which describe homogeneous but anisotropic
Kantowski-Sachs cosmological model. To describe the above
mentioned anomalies, it is shown that the anisotropic Bianchi
cosmological models are applicable by anisotropic cosmological
constant \cite{56,57} and by dark energy \cite{53,54,55,56,57,58}.
From the point of view of elementary particle physics, several
candidates have been identified for the particle carrying dark
energy or dark matter and introduced to the world of science. The
multiplicity of these candidates makes unknown origin of dark
sector of matter/energy. Because of this some scientists use other
gravitational models by regarding principle of general covariance
(laws of Physics are the same for all observers and, therefore,
must be written in terms of geometric objects), in which timelike
dynamical vector fields coupled with the geometry produce
gravitational corrections instead of the effects of unknown dark
sector. Such models are called Einstein-Aether gravity usually in
which the used timelike dynamical vector fields can be interpreted
as four vector velocity of preferred reference frames.
 In fact these dynamical timelike
vector fields break rotational symmetry of spacetime because there
is interaction between the vector field and the geometry which
will have bimetric(see introduction section of reference \cite{59}
for more discussion).

As a generalization of Einstein-Aether gravity, we consider a
scalar-vector-tensor gravity model \cite{59,60} which is made from
generalization of the well known Jordan-Brans-Dicke scalar tensor
gravity \cite{61} by transforming the background metric such that
$g_{\mu\nu}\to g_{\mu\nu}+2N_{\mu}N_{\nu}$ in which $N_{\mu}$ is
dynamical timelike four vector field. In the latter model, there is
a non-minimal interaction between the Brans Dicke scalar field and
the timelike vector field and in the present work we will see that
it is possible to have the vector field whose spatial part does
not vanish with the evolution (unless it has always been zero)
pointing towards a direction which is different to the one of the
rotational symmetry. This is what we call "preferred reference
frame effects". Several applications of the model \cite{59,60} are
studied for classical and quantum approach of FLRW cosmology
\cite{62,63,64,65,411} previously.  In the present work, we
investigate affects of a timelike dynamical massless vector field
$N_{\mu}$ interacting with the Brans Dicke scalar field, on
anisotropy property of a Bianchi I cosmology. To solve the
gravitational field equations we use dynamical system approach and
we obtain some critical manifolds with stable or quasi stable
nature in phase space for each of four directions of the vector
field which asymptotically are reduced to the de Sitter epoch with
anisotropy trajectories.
 However, we determined stability nature of the obtained critical manifolds which are dependent to spatial directions of the
 used the dynamical vector field. Organization of the paper is as follows.

In section 2, we introduce the scalar vector tensor Brans Dicke gravity model \cite{59,60} briefly.
 In section 3, we use the Bianchi I background metric to obtain exact form of dynamical field
 equations. We solve gravitational equations by using dynamical systems approach. We determine critical manifolds in phase
 space of dynamical equations in presence of small perturbations of anisotropy property of
 the spacetime for each of four directions of the timelike dynamical vector field.
 Section 4 assign to concluding remark and outlook of the work.
\section{The Model}
Let us start with the following scalar-vector-tensor-gravity
action \cite{59,60}
 \begin{equation}\label{1} I_{total}=I_{BD}+I_{N},
  \end{equation}
 where   \begin{equation}\label{2} I_{BD}=\frac{1
 }{16\pi}\int dx^4\sqrt{g}\left\{\phi R-\frac{\omega}{
 \phi}
 g^{\mu\nu}\nabla_{\mu}
 \phi\nabla_{\nu}\phi\right\},
 \end{equation}
 is the well known Brans Dicke scalar tensor
 action \cite{61},
and with definitions \begin{equation}\label{4}
F_{\mu\nu}=2(\nabla_{\mu}N_{\nu}-\nabla_{\nu}N_{\mu}),
~~~~~~~\Omega_{\mu\nu}=2(\nabla_{\mu}N_{\nu}+\nabla_{\nu}N_{\mu}),
\end{equation}
the action functional
\begin{align}\label{3}
I_N=&\frac{1}{16\pi}\int
 dx^4\sqrt{g}\{\zeta(x^{\nu})(g^{\mu\nu}N_{\mu}N_{\nu}+1)+2\phi F_{\mu\nu}F^{\mu\nu}
-U(\phi, N_{\mu})\notag\\
&-\phi N_\mu
 N^{\nu}(2F^{\mu\lambda}\Omega_{\nu\lambda}+
 F^{\mu\lambda}F_{\nu\lambda}+\Omega^{\mu\lambda}\Omega_{\nu\lambda}
-
2R^{\mu}_{\nu}+\frac{2\omega}{\phi^2}\nabla^{\mu}\phi\nabla_{\nu}\phi)\},
\end{align}
describes dynamics of a unit timelike four vector field $N_\mu$
which can be considered as four velocity of a dynamical preferred
reference frame which  is couple non minimally with the Brans
Dicke scalar tensor gravity. In fact, we assume that $N_\mu$
satisfies
\begin{equation}\label{5}g_{\mu\nu}N^{\mu}N^{\nu}=-1,\end{equation} and such a model
 is named as Einstein-Aether gravity models in the literature.
 In fact, the
general covariance principle leads us to consider $N_\mu$ as a
dynamical vector field. $\zeta(x^{\mu})$ is undetermined Lagrange
multiplier and $U(\phi, N_{\mu})$ is interacting potential between
scalar and vector fields. Without the additional terms $\zeta$ and
$U$ the action functional (\ref{3}) is generated from (\ref{2}) by
regarding the metric transformation $g_{\mu\nu}\to
g_{\mu\nu}+2N_{\mu}N_{\nu}$. One can check references \cite{59,60}
to see detail of calculations. The action functional (\ref{3})
shows that the vector field $N_{\mu}$ is coupled as non-minimally
with the Brans Dicke scalar field $\phi.$ The action functional
(\ref{1}) is written in units $c=\hbar=1$ with Lorentzian
signature (-,+,+,+). The undetermined Lagrange multiplier
$\zeta(x^{\nu})$ controls $N_{\mu}$ to be unit timelike vector
field. According to the Mach`s principle \cite{61} the Brans Dicke
scalar field $\phi$ describes inverse of Newton`s gravitational
coupling parameter as $\phi(x)\sim\frac{1}{G(x)}$ and its
dimension is $(lenght)^{-2}$ in units $c=\hbar=1$. Authors in
reference \cite{61} showed that to have an inflationary model for
FRW metric the Brans Dicke scalar field should be a raising
function versus the cosmic comoving time and this theory reduces
to general theory of relativity
at $\omega\to\infty.$ While our mathematical calculations show
that the constraint on the $\omega$ parameter in Brans Dicke
gravity does not valid in our modified scalar vector model.
 This is because our stable
solutions are happened at small values of $\omega$ parameter. But
fortunately, these solutions have stable nature in phase space
with a raising function for the Brans Dicke field solution in both
inflationary epochs and for all different directions of the vector
field. With this view, it seems there is non minimal interaction
between the timelike vector field and the Brans Dicke scalar field
can resolve the inconsistency problem (why constraint on large
values of the $\omega$ parameter does not consistent with our
modified gravity theory). $g$ is absolute value of determinant of
the metric field $g_{\mu\nu}$. Present limits of dimensionless
Brans Dicke $\omega$ parameter  based on time-delay experiments
\cite{66,67,68,69} requires $\omega\geq4\times10^{4}$, but we will
see  this constraint can be violated in the alternative gravity
model (\ref{1}). Recently, authors of the work \cite{Hash}
investigated on constraining an exact Brans Dicke gravity with
recent observations and obtained $\omega>1627.$

By varying (\ref{1}) with respect to $\zeta(x^{\nu})$ we obtain
(\ref{5}) and by varying (\ref{1}) with respect to  the fields
$\phi,$ $N^{\mu}$ and $g^{\mu\nu}$, we obtain corresponding
dynamical equations respectively as follows
\begin{align}\label{phi1}
&\frac{2\omega\Box\phi}{\phi}-\frac{\omega
g^{\mu\nu}\partial_{\mu}\phi\partial_{\nu}\phi}{\phi^2}-\frac{4\omega
N^{\mu}N^{\nu}\partial_{\mu}(\sqrt{g}\partial_{\nu}\phi)}{\phi\sqrt{g}}-\frac{\partial
U(\phi, N_\mu,)}{\partial
\phi}\notag \\
&-\frac{4\omega\partial_{\mu}(N^{\mu}N^{\nu})
\partial_{\nu}\phi}{\phi}
-\frac{4\omega\Gamma^{\mu}_{\mu\alpha}N^{\alpha}N^{\nu}\partial_{\nu}\phi}{\phi}
-\frac{4\omega\Gamma^{\nu}_{\mu\lambda}N^{\mu}N^{\lambda}\partial_{\nu}\phi}{\phi}
+\frac{2\omega N^{\mu}N^{\nu}\partial_{\mu}\phi\partial_{\nu}\phi}{\phi^2}\notag\\
&+R-2N^{\mu}N^{\nu}R_{\mu\nu}+2F_{\mu\nu}F^{\mu\nu}
-N_{\mu}N^{\nu}
\{2F^{\mu\lambda}\Omega_{\nu\lambda}+F^{\mu\lambda}F_{\nu\lambda}+
\Omega^{\mu\lambda}\Omega_{\nu\lambda}\}=0,
\end{align}
\begin{align}\label{vec}
&\frac{[4F_{\mu\nu}-N_{\mu}N^{\lambda}(F_{\lambda\nu}+3\Omega_{\lambda\nu})+N_{\nu}N^{\lambda}(F_{\lambda\mu}-\Omega_{\mu\lambda}
)]\partial^{\mu}(\sqrt{g}\phi)}{\sqrt{g}\phi}\notag \\
&-\frac{\partial
U(\phi, N_\mu,)}{\phi\partial N^\mu}
+\nabla^{\mu}[4F_{\mu\nu}-N_{\mu}N^{\lambda}(F_{\lambda\nu}+3\Omega_{\lambda\nu})+N_{\nu}N^{\lambda}(F_{\lambda\mu}-\Omega_{\mu\lambda})]\notag \\
&+N_{\mu}(F_{\nu\lambda}+3\Omega_{\nu\lambda})\nabla^{\mu}N^{\lambda}+N^{\lambda}
(F_{\lambda\mu}+3\Omega_{\lambda\mu})\nabla_{\nu}N^{\mu}-N_{\lambda}(F_{\nu\mu}-\Omega_{\mu\nu})
\nabla^{\mu}N^{\lambda}\notag \\
&-N^{\lambda}(F_{\lambda\mu}-\Omega_{\mu\lambda})\nabla^{\mu}N_{\nu}+2N^{\mu}R_{\mu\nu}-\frac{2\omega
N^{\mu}\partial_{\mu}\phi\partial_{\nu}\phi}
{\phi^2}-\frac{\zeta(x^{\alpha})N_{\nu}}{\phi}=0,
\end{align}
and
\begin{align}\label{Ein1}
&G_{\mu\nu}=\frac{8\pi}{\phi}T^{matter}_{\mu\nu}+\frac{\omega\partial_{\mu}\phi\partial_{\nu}\phi}{\phi^2}+
\frac{\partial_{\mu}(\sqrt{g}\partial_{\nu}\phi)}{\sqrt{g}\phi}
-\frac{\zeta(x^{\alpha})N_{\mu}N_{\nu}}{\phi}\notag \\
&+\frac{2\Box(\phi N_{\mu}N_{\nu})}{\phi}-\frac{g_{\mu\nu}}{2\phi}\{2\Box\phi+\frac{\omega
g^{\alpha\beta}\partial_{\alpha}\phi\partial_{\beta}\phi}{\phi}
-2\phi F_{\alpha\beta}F^{\alpha\beta}+2\phi
N_{\alpha}N^{\beta}F^{\alpha\lambda}\Omega_{\beta\lambda}\notag \\
&+\phi
N_{\alpha}N^{\beta}(F^{\alpha\lambda}F_{\beta\lambda}+\Omega^{\alpha\lambda}\Omega_{\beta\lambda})
+2N^{\alpha}N^{\beta}(\phi R_{\alpha\beta}
-\frac{\omega\partial_{\alpha}\phi\partial_{\beta}\phi}{\phi})\}+
\frac{U(\phi,N_{\mu})}{\phi} g_{\mu\nu},
\end{align}
 where
$$\Box=\frac{1}{\sqrt{g}}\partial_{\mu}(\sqrt{g}g^{\mu\nu}\partial_{\nu}).$$
We now set the above dynamical equations for anisotropic Bianchi I
cosmological model as follows.
\section{Bianchi I cosmology}
Spatially homogeneous but anisotropic dynamical flat universe
given by the Bianchi I metric has the following line element from
point of view of free falling comoving observer \cite{40}.
 \begin{equation} \label{Bian}ds^2=-dt^2+e^{2a(t)}\{e^{-4b(t)}d\mathbf{x}^2+e^{2b(t)}(d\mathbf{y}^2+d\mathbf{z}^2)\},
 \end{equation}
where ${\mathbf{x},\mathbf{y},\mathbf{z}}$ are Cartesian spatial
coordinates of the comoving observer and $t$ is cosmic time. In
the metric equation (\ref{Bian}), we assume that the spatial parts
have a cylindrical symmetry for which $e^{a(t)}$ is  global
isotropic scale factor and $b(t)$ represents deviations from the
isotropy.
 Substituting (\ref{Bian}) the equation (\ref{5}) reads $1=N^{t2}(t)-e^{2a-4b}N^{x2}(t)-e^{2(a+b)}[N^{y2}(t)+N^{z2}(t)]$
in which time dependent components of the vector field $N_{\mu}$
 should satisfy the following parametric identities
 \begin{equation}\label{7} N_{\mu}(t)=\left(%
\begin{array}{c}
  N_t \\
  N_\mathbf{x} \\
  N_\mathbf{y} \\
  N_\mathbf{z} \\
\end{array}%
\right)=\left(%
\begin{array}{c}
  \cosh\alpha \\
  e^{a-2b}\sinh\alpha\cos\beta \\
  e^{a+b}\sinh\alpha\sin\beta\cos\gamma \\
  e^{a+b}\sinh\alpha\sin\beta\sin\gamma \\
\end{array}%
\right),\end{equation} where the parameters $(\alpha,\beta,\gamma)$
denote polar directions of the vector field $N_{\mu}.$
  Substituting (\ref{7}) one can calculate $F_{\mu\nu}$ and $\Omega_{\mu\nu}$ as follows
  \begin{equation}\label{8}F_{t\mathbf{x}}=2(\dot{a}-2\dot{b})N_\mathbf{x},~~~F_{t\mathbf{y}}=2(\dot{a}+
  \dot{b})N_\mathbf{y},~~~F_{t\mathbf{z}}=2(\dot{a}+\dot{b})N_\mathbf{z},
  \end{equation}
  and
  \begin{align}\label{9}
  &\Omega_{t\mathbf{x}}=-2(\dot{a}-2\dot{b})N_\mathbf{x},
&&  \Omega_{t\mathbf{y}}=-2(\dot{a}+\dot{b})N_\mathbf{y},
&& \Omega_{t\mathbf{z}}=-2(\dot{a}+\dot{b})N_\mathbf{z},\\
&\Omega_{xx}=-4(\dot{a}-2\dot{b})e^{2a-4b}N_t,&&
\Omega_{yy}=\Omega_{zz}=-4(\dot{a}+\dot{b})e^{2a+2b}N_t.\notag
  \end{align}
 To solve the dynamical field
equations for the line element (\ref{Bian}), we remember symmetry
property of the Einstein`s tensor in left hand side of the metric
equation (\ref{Ein1}) where all non diagonal components have zero
values and its diagonal components are
\begin{align}
&G^{t}_t=3(\dot{a}^2-\dot{b}^2),\\
&G^x_x=2\ddot{a}+2\ddot{b}+3\dot{a}^2+6\dot{a}\dot{b}+3\dot{b}^2,
\end{align}
and
\begin{equation}
G^y_y=G^z_z=2\ddot{a}-\ddot{b}+3\dot{a}^2-3\dot{a}\dot{b}+3\dot{b}^2,
\end{equation}
and so non diagonal components in
right side of the metric equation (\ref{Ein1}) should be set with
zero values.
  This is done by choosing some different ansatz for direction of
  the vector field $N_\mu(t)$ such that $(N_t\neq0,N_{x,y,z}=0),$
$(N_{x}\neq0,N_{t,y,z}=0)$ and $(N_{y}\neq0,N_{t,x,z}=0).$ Without
to use the above ansatz there is an inconsistency between right
and left hand sides of the metric equation (\ref{Ein1}). Hence, we
solve dynamical field equations
  separately for each of the above mentioned choices as follows.
  \subsection{Metric solution for $N_t\neq0,N_{x,y,z}=0$} In this case, we must be set $\alpha=0$ in the equation
   (\ref{7}) for which we will have
   $$N_t=1,\qquad N_{x,y,z}=0.$$
   In this case, one can show that the field equations (\ref{phi1}), (\ref{vec}) and $G^t_{t},G^x_x,G^y_y=G^z_z$ components of the metric equation
   (\ref{Ein1}) will have the following forms respectively.
   \begin{align}
 &  \omega\frac{\dot{\phi}^2}{\phi^2}-2\omega
   \frac{\ddot{\phi}}{\phi}+2\omega\frac{\dot{a}\dot{\phi}}{\phi}-6\dot{a}^2-6\dot{b}^2-4\ddot{a}-\frac{1}{3}\frac{\partial U(\phi)}{\partial\phi}=0,\label{phi2} \\
 &\frac{\zeta(t)}{\phi(t)}=2\omega\frac{\dot{\phi}^2}{\phi^2}-6\dot{a}^2-12\dot{b}^2-6\ddot{a},\\
& 2\frac{\ddot{\phi}}{\phi}+\frac{5\omega}{2}\frac{\dot{\phi}^2}{\phi^2}+6\frac{\dot{a}\dot{\phi}}{\phi}-12\dot{a}^2-15
   \dot{b}^2-9\ddot{a}+\frac{U(\phi)}{\phi}=0,\label{Gttt}\\
   &\frac{\ddot{\phi}}{\phi}-4\ddot{a}-6\dot{a}^2-6\dot{b}^2+\frac{3\dot{a}\dot{\phi}}{\phi}-\frac{\omega}{2}
   \frac{\dot{\phi}^2}{\phi^2}=0, \label{Gxxt}\\
&\ddot{b}+3\dot{a}\dot{b}=0,\qquad \dot{b}=Ke^{-3a},\label{Gyyt}
\end{align}
   in which we substitute $T^{matter}_{\mu\nu}=0,$ because the model under consideration is a creative matter alternative gravity model in which the Brans Dicke scalar field $\phi$ and the vector field $N_{\mu}$ play the role of the matter and we define the integral constant $K=\dot{b}(t)_{a=0}$ to be
   initial velocity of anisotropy at primordial inflation $(a=0)$.  Because in the primordial inflation size of the space time has smallest
   scale and the cosmic system is in a high energy state for which we can consider that the background metric is flat Minkowski
   at the primordial inflation $a(0)\approx0.$
   The equation (\ref{Gyyt})
   shows that at late inflationary period where $a>>1$ the
   anisotropy expansion velocity $\dot{b}$ can be negligible and so one can infer that the anisotropy property of this space time is vanishing by the expansion such that
   \begin{equation}\lim_{a\to+\infty}\bigg(\frac{\dot{b}}{\dot{a}}\bigg)=\lim_{a\to+\infty}\bigg(\frac{db}{da}\bigg)=\lim_{a\to+\infty}Ke^{-3a}=0.\end{equation}
   Furthermore, in this case we set $U(\phi,N_t=1)\equiv
   U(\phi).$
 To study chaotic inflation in the well known alternative scalar tensor gravity theories,
   one usually use power law self interaction potential for the inflaton field as $U(\phi)\sim \phi^n$ (see for instance
   \cite{2006} and references therein) but we feel that a simpler linear form $n=1$ may be enough to study inflation phase of
   the model under consideration because this model is two fluids model for which the second field is non-minimal time like vector
   field $N_{\mu}$
   which should support the inflation. Hence, we check just linear potential in the present work and some complicated forms $n\neq1$
   dedicated to our future works.
    However, by defining
    \begin{equation}\label{difNt}\frac{\dot{\phi}}{\phi}=\psi(t),~~~\dot{a}=H(t),~~~U(\phi)=C\phi,
    \end{equation}
the equations (\ref{phi2}), (\ref{Gttt}) and (\ref{Gxxt}) read
\begin{align}
&\dot{H}=-3H^2+\frac{3\omega
H\psi}{2(1+2\omega)}-\frac{\omega(6+15\omega)}{2(1+2\omega)}\psi^2-\bigg(\frac{5+8\omega}{1+2\omega}\bigg)\frac{C}{2},\label{Htt}\\
&\dot{\psi}=-\bigg(\frac{2+\omega}{1+2\omega}\bigg)\frac{\psi^2}{2}+\bigg(\frac{3-2\omega}{1+2\omega}
\bigg)H\psi-\frac{C}{2(1+2\omega)},\label{psitt}
\end{align}
and
\begin{equation}
\dot{b}=Ke^{-3a}=\sqrt{H^2+\frac{\omega(13+29\omega)}{6(1+2\omega)}\psi^2
-\frac{4\omega H\psi}{3(1+2\omega)}+
\bigg(\frac{9+16\omega}{1+2\omega
}\bigg)\frac{C}{12}}.
\end{equation}
 We obtain solutions of the nonlinear differential equations (\ref{Htt}) and (\ref{psitt}) near its critical manifolds and
 study stability conditions of the obtained solutions. Critical manifolds in phase space $\{H,\psi\}$ are obtained by solving the equations
 $\dot{H}=0=\dot{\psi}$ for which the above dynamical equations reduce to the following relations.
\begin{align}
&\frac{H_c}{\psi_c}=y_c,\label{Hct}\\
&y_c^{\pm}(\omega)=\frac{(32\omega^2-25\omega-30)}{12(1+2\omega)}\pm\frac{\sqrt{1024\omega^4-1936\omega^3-
743\omega^2+2340\omega+1140}}{12(1+2\omega)},\label{yct}\\
&\psi^{\pm}_c(\omega)=\sqrt{\frac{-C}{2y_c(2\omega-3)+2+\omega}},\label{psic}
\end{align}
and
\begin{equation}\label{bdotNt}\dot{b}^{\pm}_c(\omega)=Ke^{-3a_c}=\sqrt{\psi_c^2\bigg[y_c^2-\frac{4\omega
y_c}{3(1+2\omega)}+\frac{\omega(13+29\omega)
}{6(1+2\omega)}\bigg]+\frac{C}{12}\bigg(\frac{9+16\omega}{1+2\omega}\bigg)},\end{equation}
which at large values of the $\omega$ parameter asymptote to the
following forms respectively.
\begin{align}
&y_c^+\approx2.7\omega-3.6-\frac{0.5}{\omega}+O(\omega^{-2}),\\
&y_c^-\approx0.2-\frac{0.3}{\omega}+O(\omega^{-2}),\\
&\dot{b}_c^+\approx\sqrt{\frac{-123C}{\omega}}\bigg[0.04+\frac{0.03}{\omega}+O(\omega^{-2})\bigg],\label{exp1}\\
&\dot{b}_c^-\approx\sqrt{-35C}\bigg[0.13+\frac{0.02}{\omega}+O(\omega^{-2})\bigg],\label{exp2}\\
&\psi_c^+\approx\sqrt{\frac{-6C}{\omega}}\bigg[0.13+\frac{0.17}{\omega}+O(\omega^{-2})\bigg],\label{exp3}\\
&\psi_c^-\approx\sqrt{\frac{-30C}{\omega}}\bigg[0.13+\frac{0.02}{\omega}+O(\omega^{-2})\bigg]\label{exp4}.
\end{align}
The above relations show that to have real fields, we should
choose $C>0$ in limits $\omega\to\infty$ (see Eq.33) but for small
regions of
 the $\omega$ parameter the case $C<0$ has still some acceptable solutions (see diagrams 2-a and 2-b).
  We plot diagrams of the critical solutions (\ref{Hct}), (\ref{yct}), (\ref{psic}) and (\ref{bdotNt}) versus
  the $\omega$ parameter in Figures 1 and 2 for $C=\pm1$. In fact $C>0$ describes a repeller Brans Dicke potential while
   $C<0$ shows an attractor potential. Diagram of 1-a shows all possible real values for $y_c^{\pm}.$ Diagram of 1-b shows
   $\dot{b}_c$ versus $\omega$ parameter at critical manifolds $y_c^\pm.$ Diagram of the Figure 1-c shows variation of the $\psi_c$
    versus $\omega$ parameter at points $y_c^\pm.$ However, time trajectories of the fields $H(t)$ and $\psi(t)$ are obtained by
    the linearized form of the equations (\ref{Ht}) and (\ref{psit}) near the critical values $(H_c,\psi_c)$ as follows
    \begin{equation}\label{Jac}\frac{d}{dt}\left(%
\begin{array}{c}
  H(t) \\
 \psi(t) \\
\end{array}%
\right)=\left(%
\begin{array}{cc}
  J_{11} & J_{12} \\
  J_{21} & J_{22} \\
\end{array}%
\right)\left(%
\begin{array}{c}
  H(t) \\
  \psi(t) \\
\end{array}%
\right),\end{equation} where the constants $J_{ij}=\frac{\partial
\chi_{i}}{\partial q_j}$ with $\chi_i=\{\dot{H},\dot{\psi}\}$ and
$q_j=\{H,\psi\}$  are components of the Jacobi matrix of two
dimensional phase space $(H,\psi).$ They are calculated at the
critical point $y_c$ by applying (\ref{Ht}) and (\ref{psit}) as
follows
\begin{equation}J_{ij}=\frac{\psi_c}{2(1+2\omega)}\left(%
\begin{array}{cc}
 3\omega-6(1+2\omega)y_c& 3\omega y_c-2\omega(15\omega+6)\\
 2(3-2\omega)  & 2(3-2\omega)y_c-2(2+\omega) \\
\end{array}%
\right),\end{equation} which its secular equation
$\det(J_{ij}-\sigma\delta_{ij})=0$ reads
\begin{align}\label{sigm}
&(2+4\omega)^2\sigma^2+
2(1+2\omega)[6(1+2\omega)\psi_c+2(2\omega-3)y_c+4-\omega]\sigma \notag \\
&+\psi_c[(48\omega^2-48\omega-36)y_c+24\omega^2+60\omega+24]-120\omega^3+126\omega^2+60\omega=0.
\end{align} This secular
equation has two different solutions versus the $\omega$ parameter
and they occur at Figures 1-d and 2-c for $C\geq0$ and $C<0$
respectively. In fact for $C\geq0$ vertical axis in Figure 1-d
shows real part of complex eigenvalues whose negative numeric
values describe spiral stable nature for the system while negative
numeric values in the vertical axis of the figure 2-c show stable
nature for the system. These diagrams show just one negative
numeric value for the eigenvalues for large $\omega$ and the
second eigenvalue has not negative numeric value. This means that
the system will be quasi stable for $\omega>>1.$ This can be
follow by asymptotic behavior of the eigenvalues at
$\omega\to\infty$ which are obtained from (\ref{sigm}) as follows
\begin{equation}
\sigma_{1}^\pm\approx0.016+2.74\sqrt{\omega}+O(\omega^{-1/2})>0,~~~for~~~y_c^-\end{equation}
and
\begin{equation}
\sigma_{2}^\pm\approx0.016-2.74\sqrt{\omega}+O(\omega^{-1/2})>0,~~~for~~~y_c^+.\end{equation}
Diagrams of the eigenvalues in Figures 1-d and 2-c show stable
nature for our parametric solutions just for numeric values of the
$\omega$ parameter where all two eigenvalues take on negative
numeric values. This  occurs at small $\omega.$ For instance for
 ansatz $\omega=0.7$ with $C=+1$ we
 obtain numeric values for the critical points and corresponding
 eigenvalues as $y_c^+=0.45$ with $\sigma_1^+=-0.6+0.8i$ and $\sigma_2^+=0.2-3.5i$ and
$y_c^-=-2.7$ with eigenvalues $\sigma_1^-=-0.7-1.7i$ and
$\sigma_2^-=-1.8+0.9i.$ These critical points have $\psi^+_c=0.9i$
and $\psi_c^-=0.3i$ and so are not physical solutions because
$\psi_c^{\pm}$ have imaginary numeric value and are not real
fields. While one can check numeric values of the critical points
for $\omega=0.7$ with attractor potential $C=-1$ to be physical as
$y_c^+=0.5$ with corresponding values for $\psi^+_c=0.9$ and
$\sigma_1^+=-1.5+1.6i$ and $\sigma_2^+=-1.5-1.6i$ which describes
a spiral stable physical solution. And for second critical point
$y_c^-=-2.7$ with $\psi_c^-=0.3$ and $\sigma_1^-=-1.7+1.6i$ and
$\sigma_2^-=-1.7-1.6i$ which describes a spiral stable physical
solution also. By looking at the series forms
(\ref{exp1}),(\ref{exp2}) and (\ref{exp4}) we must be choose
$\omega<0(>0)$ for the case $C>0(<0)$ but (\ref{exp3}) shows that
for physical  real fields we  must be set just $C<0$ and so we
collect some physical numeric solutions for the critical points
and corresponding eigenvalues in the table 1 for different
$\omega$ values. In the last column of the table we call stability
nature of the physical (means with real fields for $y_c$ and
$\psi_c$) solutions. To obtain numerical solutions given in the
table 1 we set ansatz $C=-1.$
\begin{table}[htp]
\begin{center}
\caption{critical points, eigenvalues and their nature for $C=-1$}
\begin{tabular}{|c|c|c|c|c|c|}
  \hline
  $\omega$ & $(y_c^+,y_c^-)$ & $(\psi_c^+,\psi_c^-)$ & $(\sigma_1^+,\sigma_2^+)$ & $(\sigma_1^-,\sigma_2^-)$ & $Nature$ \\
  \hline
 $+\infty$ & $(+\infty,0)$ & $(0,0)$ & $(+\infty,+\infty)$ & $(-\infty,-\infty)$ & quasi~stable \\
 +100 & $(+263.02,-0.22)$ & $(+0.003,+0.073)$ & $(+2.88,-260.37)$ & $(+29.64,-29,83)$ & quasi~stable \\
  +0.9 & $(+0.5,-2.1)$ & $(+0.8,+0.4)$ & $-1.3\pm1.5I$ & $-1.3\pm1.6I$ & spiral~stable \\
  +0.6 & $(+0.4,-2.7)$ & $(+0.96,+0.27)$ & $-1.7\pm1.6I$ & $-2\pm1.3I$ & spiral~stable \\
  +0.5 & $(+0.4,-3.3)$ & $(+1.1,+0.25)$ & $-1.8\pm1.5I$ & $-2.5\pm0.56I$ & spiral~stable \\
  +0.1 & $(+0.3,-4.8)$ & $(+8.7,+0.16)$ & $(-1.3,-5.6)$ &  $(-0.67,-26.62)$ & stable \\
  0.0 & $(+0.31,-5.3)$ & $(+2.9,+0.17)$ & $(-1,-8.7)$ & $(-17.94,-0.52)$ & stable \\
  -0.1 & $(+0.3,-6)$ & $(+8.7,+0.16)$ & $(-1.12,-26.34)$ & (-0.4,-26.5) & stable \\
  -0.5 & $\infty$ & $\infty$ & $\infty$ & $\infty$ & undetermined \\
  -0.6 & $(+0.5,-0.4)$ & $(+0.2,+3.3)$ & $9.1\pm10.8I$ & $(-4.1-0.7I,-53.4+0.1I)$ & spiral~stable \\
  -0.9 & $(-0.4,+0.45)$ & $(+0.15,+0.56I)$ & $(+26.97,+1.51)$ &  $(+0.36-8.2I,+0.02+6.5I)$ & un~physical\\
   -100 & $(-270.3,0.22)$ & $(+0.003,+0.07)$ & $(+273.2,+2.8)$ & $+0.02\pm29.5I$ & un~stable \\
 $-\infty$ & $(-\infty,0)$ & $(0,0)$ & $(+\infty I,+\infty I)$ & $(-\infty I,-\infty I)$ & spiral~quasi~stable \\
  \hline
\end{tabular}
\end{center}
\end{table}
In fact, $y_c^{\pm}<0(>0)$ in the table 1 describe a collapsing
(expanding) universe because physical values for critical time
fluctuations of the Brans Dicke field $\psi_c$ should be take on
some positive values and so negative numeric values for $y_c$
readds to a negative hubble parameter $H_c<0$ which describes a
collapsing universe and so in the table 1 we should consider some
real positive values for the fields $y_c^{\pm}>0$ and
$\psi_c^{\pm}>0$ which are defined inflationary expanding
universe. In fact $y_c^{\pm}>0$ given in the table 1 corresponds
to a physical stable solution because both of
   corresponding eigenvalues take on negative numeric values. By looking at this,  one can infer that for stable critical point
   the time trajectories of the Brans Dicke field should be a raising function which satisfies with physical situations. Because the Brans
   Dicke scalar field is inverse of the Newton`s gravity coupling parameter
   by regarding the Mach`s principle (see \cite{61}). Also, one can see that for $C\geq0$ the secular equation
    (\ref{sigm}) give us a complex parametric eigenvalues
   and so to plot possible acceptable numeric values for the eigenvalues we must to plot real part of the complex parametric $\sigma(\omega)$ which are
   appeared in Figure 1-d. In the latter case one call spiral stable nature for the dynamical system under consideration with $Re(\sigma)<0$.
   According to the
dynamical
  system approach
   (see \cite{62}), we know that all possible stable solutions
  should have negative numeric values for the real eigenvalues and spiral stable nature for the solutions
  if the eigenvalues become complex number
  with negative numeric values for its real part. We plotted arrow diagrams of the dynamical field
  equations given by (\ref{Htt}) and (\ref{psitt}) in Figure 3 for choices $\omega=0$ and $\omega=\pm0.1$
  given by the table 1. These arrow diagrams show stable point where the arrows approach to a fixed point finally.
  This means stability of the metric solutions in studying of
  the dynamical system approach which show sink hole for
  $(H_c,\psi_c)>0.$
    This is appeared for attractor potential $C=-1.$
   However, one can solve (\ref{Jac}) to obtain time trajectories of
the fields $H(t)$ and $\psi(t)$ around the stable nature critical
points $(H_c,\psi_c)$ given by the table 1 such that
\begin{equation}\label{Ht}H(t)=H_c+O_{H}(e^{\psi_c\sigma_1t}-e^{\psi_c\sigma_2t}),\end{equation}
and
\begin{equation}\label{psit}\psi(t)=\psi_c+O_{\psi}(e^{\psi_c\sigma_1t}-e^{\psi_c\sigma_2t}),\end{equation}
in which $O_{H,\psi}$ are integral constants and they should be
fixed with initial conditions on the system.  In the above
solutions, we choose origin of the time to be critical time
$t_c=0$ for which $H(0)=H_c$ and $\psi(0)=\psi_c.$ Also, numeric
values of the critical fields $H_c$ and $\psi_c$ for the stable
critical points $y_c$ should be substituted by the numeric values
in the table 1. By integrating the Equations (\ref{Ht}) and
(\ref{psit}),  we obtain
\begin{equation}\label{aT}a(t)=a_c+H_ct+\frac{O_H}{\psi_c\sigma_1}(e^{\psi_c\sigma_1t}-1)-\frac{O_H}{\psi_c\sigma_2}(e^{\psi_c\sigma_2
t}-1),
\end{equation} and
\begin{equation}\label{phiT}\ln\bigg(\frac{\phi(t)}{\phi_c}\bigg)=\psi_ct+\frac{O_\psi}{\psi_c\sigma_1}(e^{\psi_c\sigma_1t}-1)-\frac{O_\psi}{\psi_c\sigma_2}(e^{\psi_c\sigma_2 t}-1),
\end{equation} where we assumed
\begin{equation}a(0)=a_c,~~~\phi(0)=\phi_c.\end{equation} We can
substitute (\ref{Ht}) into the definition of the deceleration
parameter $-q=1+\frac{\dot{H}}{H^2}$ to obtain
\begin{equation}\label{qT}q(t)\approx-1+\frac{\psi_c\sigma_{1}}{O_H}e^{-\psi_c\sigma_1t}\end{equation}
 for $\sigma_1>\sigma_2$ and vice versa. Acceleration of
 the universe say us $q<0$ for both primordial inflation or late time inflation. At late time
 inflation (the present epoch of the anisotropic universe),
 the above time dependent deceleration parameter reaches to the following limit \begin{equation}\label{qlim}\lim_{t\to\infty}q(t)=-1,
 \end{equation}
and for beginning of the
primordial inflation we have
\begin{equation}\lim_{t\to-\infty}q(t)\approx -1,
\end{equation} while for
end of primordial inflation we
can evaluate
\begin{equation}\label{q}q(0)=-1+\frac{\psi_c\sigma_{1}}{O_H}<0,\end{equation} which reads
\begin{equation}\label{OH}O_H>\psi_c\sigma_{1}
\end{equation} if $\sigma_1>\sigma_2$ and vice versa.
In this view, we assumed that the primordial inflation is begin at
$t\to-\infty$ for which isotropic part of space time scale factor
vanishes $a(-\infty)=-\infty$ and it is ended at the time $t=0$
where at this duration of expansion the declaration parameter
takes on some negative numeric values. By substituting (\ref{aT})
 the equation $\dot{b}=Ke^{-3a}$ reads
\begin{equation}\label{bT}\dot{b}(t)=K\exp\{-8a_c-8H_ct-\frac{8O_H}{\psi_c\sigma_1}(e^{\psi_c\sigma_1t}-1)+\frac{8O_H}{\psi_c\sigma_2}(e^{\psi_c\sigma_2 t}-1)\},
\end{equation}
 in which for stable physical solutions we have $\sigma_{1,2}<0$ and $\psi_c>0$ and so the above anisotropy velocity decreases with $t\to\infty.$
  By regarding the condition (\ref{OH}) the solutions (\ref{bT}) and (\ref{Ht}) read to the following forms for
  numeric values of the critical point $\omega=0.1$ given in the table 1.
  \begin{equation}\label{bdt}\dot{b}(t)=\exp\{-20.88-16\exp{(-11.31t)}+16\exp{(-48.72t)}\}
\end{equation} and \begin{equation}\label{adt}\dot{a}(t)=H(t)=2.61-22.62\exp{(-11.31t)}+22.62\exp{(-48.72t)}\end{equation}
where we set
\begin{align}\label{asym}
&K=e^{8a_c}, &&O_H=2\psi_c\sigma_1.
\end{align}
Diagrams of these solutions given in the Figure 4 show that
velocity of anisotropy time trajectory of the space time
$\dot{b}(t)$ is smaller than the Hubble parameter time trajectory
(velocity of isotropy part of the space time) by raising the
cosmic time but it dose not never vanishes. Now, we investigate
some suitable conditions for the solutions (\ref{bT}) and
(\ref{aT}) such that they can be satisfy both of primordial or
late time inflations for line element (\ref{Bian}).
   \subsubsection{Primordial inflation} In the primordial inflation, the scale factor of the
   space time (\ref{Bian}) takes on smallest size and the system is in high energy state.
   Thus, one can assume that the background metric is flat Minkowski
  at duration of this phase of the expansion and so
  we should substitute boundary conditions $a(0)=a_c=0$ at end of the primordial inflation in the equations (\ref{asym})
  such that
\begin{align}
&K=1, \\
&b(t)\approx-\frac{e^{-1761(t-t_p)}}{1761},\label{bT0}
\end{align}and
\begin{equation}\label{aTP}a(t)\approx32.13(t-t_p)+14.76(t-t_p)^2,\end{equation}
where $t_p=-0.01919308906$ is particular cosmic time for which
$\dot{b}(t_p)=1$ and to calculate the above integral solutions we
use first order Taylor series expansion in the exponent of the
functions $\dot{b}$ and $\dot{a}$ given by the equations
(\ref{bdt}) and (\ref{adt}). The solution (\ref{bT0}) shows that
for beginning of the primordial inflation $(t<<t_p)$ the
anisotropy of the space time is not negligible but is for late
time inflationary phase of the space time expansion $(t>>t_p)$. By
substituting the solutions (\ref{bT0}) and (\ref{aTP}) into the
definitions of the metric fields we obtain
 \begin{equation}\label{b0}g_{xx}=\exp\{64.26(t-t_p)^2+29.52(t-t_p)^2+0.0023\exp[-1761(t-t_p)]\}
  \end{equation}
and
  \begin{equation}\label{Sigma}g_{yy}=g_{zz}=\exp\{64.26(t-t_p)^2+29.52(t-t_p)^2-0.001\exp[-1761(t-t_0)]\}.\end{equation}
Diagrams of the solutions (\ref{bdt}) and (\ref{adt}) are plotted
in Figures 4-a and 4-b. Diagrams of the equations (\ref{bT0}) and
(\ref{aTP}) are plotted in Figures 4-c and corresponding metric
components of line element (\ref{Bian}) are plotted in Figures
4-d. They show isotropic scale factor raises faster than the
anisotropic scale factor and at duration of the inflation there is
not more different time trajectories between $g_{xx}$ and
$g_{yy}=g_{zz}$. For this stable solution, we obtain matter
density, directional pressures and corresponding barotropic
indexes which are defined by the following formulas and whose
diagrams are plotted in Figures 5.
\begin{align}
&\frac{\rho(T)}{\rho_c}=G^t_t=3(a^{\prime2}-b^{\prime2}),\\
&\frac{p_x}{\rho_c}=2a^{\prime\prime}+2b^{\prime\prime}+3a^{\prime2}+6a^{\prime}b^{\prime}+3b^{\prime2},
\label{px} \\
&\frac{p_y}{\rho_c}=2a^{\prime\prime}-b^{\prime\prime}+3a^{\prime2}-3a^{\prime}
b^{\prime}+3b^{\prime2},\label{py}
\end{align}
in which
\begin{equation}\prime\equiv\frac{d}{dT},~~~T=t\sqrt{\rho_c},~~~\rho_c=3(H_c^2-\dot{b}_c^2),
\end{equation}
and barotropic indexes are obtained by
\begin{equation}\gamma_x=\frac{p_x}{\rho},~~~\gamma_y=\frac{p_y}{\rho},~~~\bar{\gamma}=\frac{\gamma_x+2\gamma_y}{3},
\end{equation}
where explicit form of time dependence of the above functions are not shown because they have long length.
  They do not show negative pressures for after the primordial
inflation $t>0$ which means that the Brans Dicke scalar vector
matter behaves as regular visible matter instead of dark sector
for support of spacetime inflation. In fact the Figure 5-d shows
that before end of primordial inflation $t\to0$ we have
$\gamma_{x,y}<0$ which means that the matter field of the model
behaves as dark sector. Red line in the Figure 5-a shows the
density of the matter has maximum point firstly and then reaches
to a local minimum which means that in the end of the primordial
inflation the system reaches to the reheating phase. But with the
passage of time, it has tended to a very low density, which can be
the average density of the current phase of the world by setting
the observational data.Diagrams of 5-a and 5-b and 5-c show that
anisotropy of the space time is negligible after end of the
primordial inflation but not before than. In our calculations, we
set the dimensionless parameters so that the end time of
primordial inflation is zero. Now, in the next subsection we check
the above general solutions for the late inflationary period.
\subsubsection{Late inflationary period}
In this approach, the spacetime scale is large and the anisotropic
part of scale factor is negligible. In other word, the universe is
old (the present state) and so, we can use the asymptotic
solutions given by (\ref{aT}) such that for $t>>1$ we can write
\begin{equation}a(t)\approx \ln N+H_ct,~~~b(t)\approx0,
\end{equation}
in which we defined e-folding parameter as
\begin{equation}\label{NK}a_c=\ln N>>1.\end{equation}
In this regime, the line element (\ref{Bian}) reaches to a flat
Robertson Walker metric with scale factor
$\exp(t\sqrt{\Lambda/3})$ in the asymptotic de Sitter epoch in
which
\begin{equation}
\Lambda=3H_c^2,
\end{equation} is effective cosmological constant which supports the exponential
expansion of the universe. To determine $N$ value given by the
identity (\ref{NK}), we should use the observational data. From
the observational point of view, the total number of e-folds, must
be larger than $N>60$, which depends on the scale of inflation and
thermal history after inflation \cite{72}. Also, for critical
density of the matter we should use
$\rho_c\approx8.5\times10^-{27}kg/m^3$ from observational data.\\
 The scalar spectral index $n_s$
is observational quantity to check validity of our solutions. With
lowest order terms it is defined by slow rolling parameters
$|\epsilon|<1$ and $|\eta|<1$ of the inflation \cite{COL} as
$n_s\simeq1-6\epsilon+ 2\eta$ where
$\epsilon=\frac{m_p^2}{16\pi}\big(\frac{\partial_\phi
U(\phi)}{U(\phi)}\big)^2$ and $\eta=\frac{\partial_\phi^2
U(\phi)}{U(\phi)}$ are defined versus the derivatives of the
inflation potential. To match the observational data,  at end of
the inflation we should have $n_s\approx1$. This will be guarantee
the generation of scale invariant scalar perturbations. In fact,
the Planck satellite full mission temperature data and a first
release of polarization data on large angular scales measure the
spectral index of curvature perturbations to be
$n_s=0.968\pm0.006$ \cite{70}. By substituting the potential
$U=C\phi$ given by (\ref{difNt}), we obtain
\begin{equation} \epsilon=\frac{m_p^2}{16\pi
\phi^2},~~~\eta=0,~~~~n_s=1-\frac{3m_p^2}{8\pi \phi^2},
\end{equation} in which $m_p$ is the Planck mass. By substituting the solution (\ref{phiT}) at
limits $t\to>>\psi_c^{-1}$ the above relations reads
\begin{equation}n_s=1-\frac{3m_p^2}{8\pi\phi_c^2}e^{-\psi_c t}\approx 1,
\end{equation} which obey observational conditions.
 In the next subsection, we investigate other possible metric solutions in case where the timelike dynamical vector field is parallel with the spatial symmetry axis of the Bianchi I line element (\ref{Bian}).
\subsection{Metric solution for $(N_{x}\neq0,N_{t,y,z}=0)$}
   By substituting $N_x\neq0,N_{t,y,z}=0,$ the equation (\ref{7}) reduces to the following conditions.
   \begin{equation}\label{conNx}
   \beta=0,~~~\alpha=\frac{i\pi}{2},~~~N_x=ie^{a-2b},
   \end{equation}
    and the fields equations $\phi$, $N_{\mu}$ and $G_{\mu\nu}$  take on the following forms respectively.
    \begin{equation}\label{phiNx}2\omega\frac{\ddot{\phi}}{\phi}+2\omega(\dot{a}+4\dot{b})\frac{\dot{\phi}}{\phi}-
    \omega\frac{\dot{\phi}^2}{\phi^2}-18\dot{a}^2-86
    \dot{b}^2+
    108\dot{a}\dot{b}+2\ddot{a}+\frac{\partial U(\phi,N_x)}{\partial\phi}=0,\end{equation}
\begin{equation}\label{zetaNx}\frac{\zeta}{\phi}=-102\dot{a}^2-20\dot{b}^2+216\dot{a}\dot{b}-8\ddot{a}+12\ddot{b}-24(\dot{a}-2\dot{b})\bigg(
\frac{\dot{\phi}}{\phi} \bigg)-\frac{1}{\phi N_x}\frac{\partial
U}{\partial N^x},
\end{equation}
and $G_t^t,G_x^x,G_y^y=G_z^z$ will
be respectively
\begin{align}
&\ddot{a}+\frac{\omega}{2}\frac{\dot{\phi}^2}{\phi^2}-10\dot{a}^2+58\dot{a}\dot{b}-69\dot{b}^2-\frac{U}{\phi}=0, \label{GttNx}\\
&\ddot{a}-10\ddot{b}+830\dot{a}^2-136\dot{a}\dot{b}+111\dot{b}^2
+9\dot{a}\bigg(\frac{\dot{\phi}}{\phi}\bigg)+\frac{\omega}{2}\frac{\dot{\phi}^2}{\phi^2}+3\frac{\ddot{\phi}}{\phi}+\frac{(U+\zeta)}{\phi}=0,\label{GxxNx}
\end{align}
   and \begin{equation}\label{GyyNx}\frac{\omega}{2}\frac{\dot{\phi}^2}{\phi^2}+\frac{\ddot{\phi}}{\phi}+
   3\frac{\dot{a}\dot{\phi}}{\phi}+67\dot{b}^2-55\dot{a}\dot{b}+16\dot{a}^2+\frac{U(\phi,N_x)}{\phi}=0.\end{equation}
  The above five equations are enough to determine all fields $a,b,\phi,\zeta, U.$
  By substituting  (\ref{conNx}) and definitions
  \begin{equation}\label{difNx} \dot{a}=H,~~~\dot{b}=B,~~~\frac{\dot{\phi}}{\phi}=\psi, \end{equation}
  we can write the following identities for potential.
  \begin{equation}\label{PotNx}
  \frac{1}{N_x}\frac{\partial U}{\partial N^x}=\frac{\partial U}{\partial a}-
  \frac{1}{2}\frac{\partial U}{\partial b}=\bigg(\frac{1}{H}-\frac{1}{2B}\bigg)\dot{U},~~~\frac{\partial U}{\partial
   \phi}=\frac{\dot{U}}{\phi\psi}.
  \end{equation} By substituting (\ref{difNx}) and (\ref{PotNx})
  one can show that the dynamical field equations (\ref{phiNx}), (\ref{zetaNx}), (\ref{GttNx}),
   (\ref{GxxNx}) and (\ref{GyyNx}) are transformed to the following first order nonlinear differential equations.
   \begin{align}
   &\dot{H}=\frac{(-\omega\phi\psi^2+138B^2\phi-116BH\phi+20H^2\phi+2U)}{2\phi},\\
   &\dot{\psi}=-
   \frac{(\omega\phi\psi^2+134B^2\phi-110BH\phi+32H^2\phi+6H\phi\psi+2\phi\psi^2-2U)}{2\phi},\\
&\dot{B}=-\frac{(3\omega\phi\psi^2+42B^2\phi+58BH\phi-1584H^2\phi-10U-2\zeta)}{20\phi},\\
&\dot{U}=\frac{2HB}{5(H-2B)}\times \\
&(2986B^2\phi-11\omega\phi\psi^2-3226BH\phi-240B\phi\psi-3842H^2\phi+120H\phi\psi+10U-
   \zeta),
   \end{align}
   and
   \begin{align}
\zeta=&\frac{5\psi[\omega^2\phi\psi^2+134B^2\omega\phi-8B\omega\phi\psi-52B^2\phi-2(\omega+1)U]}{H},\notag \\
&-\frac{5\psi(\omega^2\phi\psi^2+32H^2\omega\phi+4H\omega\phi\psi+2\omega\phi\psi^2-2H^2\phi-2(\omega+1)U)}{2B}\notag \\
&+2986B^2\phi-(885\omega\phi\psi+3226H\phi+70\phi\psi)B\notag \\
&+435H\omega\phi\psi+29\omega\psi^2\phi-3842H^2\phi+90H\psi\phi+10U.
  \end{align}
  By solving the equations $\dot{H}=0=\dot{\psi}=\dot{B}=\dot{U}$, we obtain critical
   points as follows
   \begin{equation}\label{critNx}
   H_c=2B_c,~~~\psi_c=2.2484B_c,~~~U_c=-1557B_c^2\phi_c,~~~\zeta_c=4B_c^2\phi_c,~~~\omega=-618.77,
   \end{equation}
in which critical anisotropy velocity $B_c$ is arbitrary constant
which should be determined by observational data. This shows that
in order to have at least one metric solution, we must have an
anisotropy at the critical point, that is, $B_c=\dot{b}_c\neq0.$
To determine if the above critical manifolds show stable  metric
solution, we must be determine eigenvalues of the Jacobi  matrix
of the above dynamical field equations in four dimensional phase
space $\{H,B,\psi,U\}$
 similar to previous section by solving the corresponding secular equation.
 This fourth degrees algebraic equation has a negative real root (eigenvalues) such that
  \begin{equation}\label{93}\sigma_1=\sigma_2=\sigma_3=\sigma_4=-269.1022666B_c.\end{equation}
  Its negativity shows stable (unstable) nature for our obtained metric solutions in phase space with $B_c>0(B_c<0).$
   By looking at the critical point
   (\ref{critNx}) one can infer that similar to previous section this stable solution is supported with attractor critical potential
   $U_c=-1557B_c^2\phi_c.$ In fact, in the real universe we know that the anisotropy part of the  metric field should be decreasing function and in
   large scales of the universe it is negligible. While for our solution the critical hypersurface $H_c=2B_c$ in phase space shows that anisotropy
   is not negligible and grows with half velocity of the isotropic expansion.  Furthermore, in a real universe the Brans Dicke scalar field should be
   increasing function versus the cosmic times which for our critical solution is happened for $B_c>0$ with equation (\ref{critNx}) as
    $\psi_c=2.2484B_c.$ This shows that our solution can be physical.  Because  in accordance with the Mach`s principle the authors of the work
     \cite{61} showed that the Brans Dicke gravity for a FRW cosmology predicts a raising function for the Brans Dicke scalar field which behaves as
     inverse of Newton`s gravity coupling parameter. Such a asymptotic behavior for the Brans Dicke field obeyed by our obtained solution in this
       subsection.
   However, we now obtain time trajectories of the fields near the above stable critical point versus the eigenvectors (not shown) as follows
 \begin{align}
 &H(T)\approx 2B_c+H_1\exp(-269T), \label{HNx}\\
 &\psi(T)\approx  2.2484B_c+\psi_1\exp(-269T),\label{psiNx} \\
&B(T)\approx  B_c+B_1\exp(-269T),\label{BNx}
 \end{align}
 and \begin{equation}\label{UNx}U(T)\approx\frac{41.672B_c^3\phi_c \exp(-269T)}{H_1-2B_1},\end{equation}
 where we defined dimensionless cosmic time \begin{equation}T=B_ct,\end{equation} and integral constants $H_{1}, \psi_{1}, B_{1}$
 should be determined by observational data.  By integrating the solutions (\ref{HNx}), (\ref{psiNx}) and
(\ref{BNx}) we obtain time trajectory of the metric and the Brans
Dicke scalar fields as follows
\begin{align}
&a(T)\approx a_c+2T-\frac{1}{269}\frac{H_1}{B_c}\exp(-269T),\label{aNx} \\
&\ln\bigg(\frac{\phi}{\phi_c}\bigg)\approx 2.2484T-\frac{1}{269}\frac{\psi_1}{B_c}\exp(-269T),
 \end{align}
 and
 \begin{equation}\label{bNx}b(t)\approx b_c+T-\frac{1}{269}\frac{B_1}{B_c}\exp(-269T),\end{equation}
 where $a_c,b_c,\phi_c$ are initial (critical) isotropic and anisotropic metric fields and the Brans Dicke scalar field which is given at begin
 of the inflation. For this solution the matter density $\rho=G^t_t=3(H^2-B^2)$ and directional pressures $p_x=G_x^x,~p_y=G_y^y$ read
\begin{equation}\frac{\rho(T)}{B_c^2}=9+\frac{6(2H_1-B_1)}{B_c}\exp(-269T)+\frac{3(H_1^2-B_1^2)}{B_c^2}\exp(-538T).\end{equation}
 \begin{equation}\frac{p_x}{B_c^2}=27-\frac{520(H_1+B_1)}{B_c}\exp(-269T)+3\bigg(\frac{H_1+B_1}{B_c}\bigg)^2
 \exp(-538T),\end{equation} and \begin{equation}\frac{p_y}{B_c^2}=9-269\frac{(2H_1-B_1)}{B_c}\exp(-269T)+3\bigg(\frac{H_1^2-H_1B_1+B_1^2}{B^2_c}\bigg)
 \exp(-538T).\end{equation}
To obtain admissible values of the constants ${H_1,B_1}$, we
calculate extremum point of the density and we consider this
physical fact in which the matter density with all possible forms
should have positive values for all times. To do so, we set
$\frac{d\rho}{dT}=0$ and obtain
\begin{equation}\label{Tm}T_m=-\frac{1}{269}\ln\bigg(\frac{B_c(2H_1-B_1)}{B_1^2-H_1^2}\bigg),~~~\rho_m=\rho(T_m)=\frac{3B_c^2(2B_1-H_1)^2}{B_1^2-H_1^2}.
\end{equation} For a physical stable metric solution where $B_c>0$ (see the eigenvalue) the above relations show that we should have
 \begin{equation}H_1<B_1<2H_1,~~~(H_1,B_1)>0,\end{equation} and
 \begin{equation}H_1>B_1>2H_1,~~~(H_1,B_1)<0.\end{equation}  One can infer that (\ref{Tm})
  approaches to the following limits \begin{equation}\lim_{B_1\to H_1}(T_m,\rho_m)=(-\infty,+\infty),~~~\lim_{B_1\to 2H_1}(T_m,\rho_m)=
  (+\infty,9B_c^2),
  \end{equation}
  in  which left (right) side describes
 expanding (collapsing) Bianchi cosmology. By substituting $B_1=H_1$ the mass density and the pressures for expanding phase read
 \begin{align}
 &\frac{\rho(T)}{3B_c^2}=3+2\chi e^{-269T},~~~\chi=\frac{H_1}{B_c},\\
 &\frac{p_x(T)}{B_c^2}=27- 1040\chi e^{-269T}+12\chi^2 e^{-538T},
 \end{align}
  and
  \begin{equation}\frac{p_y(T)}{B_c^2}=9-269\chi e^{-269T}+3\chi^2 e^{-538T}.
  \end{equation}
  For collapsing (reheating) phase of the Bianchi I cosmology with $B_1=2H_1,$ we can obtain time
  trajectories of mass density and directional pressures of the Bianchi I cosmology as follows
 \begin{align}
 &\frac{\rho(T)}{9B_c^2}=1-\chi^2 e^{-538T},\\
&\frac{p_x(T)}{3B_c^2}=9-520\chi e^{-269T}+9\chi^2 e^{-538T},
 \end{align}
 and \begin{equation}\frac{p_y(T)}{9B_c^2}=1+\chi^2
 e^{-538T}.\end{equation} We plot diagrams of the above functions for arbitrary values $B_c=0.01$, and $\chi=0.5,10,100$ respectively in Figure 6.
We check that diagrams of the directional barotropic indexes (not
shown) are similar to diagrams of the pressures but with re-scaled
size. Negative regime of the pressures (or corresponding
barotropic index) in the diagrams show a phase transition which
originates from anisotropy velocity $B\neq0$ while one can imagine
that it is apparently originates from unknown dark sector matter.
In other words the Brans Dicke scalar vector fluid behaves as dark
sector of the cosmic matter. In Figure 6-d, one can see that arrow
diagram of the metric fields in which arrows converge to a stable
point with a non-vanishing anisotropy velocity $B\neq0.$
\subsection{Primordial inflation} Same as previous section we know that in primordial inflation the spacetime has
smallest scale and is in high energy state such that we assume $a(0)=0=b(0)$ and then the metric solution (\ref{aNx}), (\ref{bNx}) read
\begin{equation}a_p(T)=2T+\frac{\chi}{269}(1-e^{-269T}),~~~b_p(T)=T+\frac{\chi}{269}(1-e^{-269T}),
\end{equation}
with constraint condition
\begin{equation}a_c=b_c=\frac{\chi}{269}.\end{equation}
\begin{equation}\label{pmx}g_{xx}^{(p)}(T)=\exp\{-\frac{2\chi}{269}(1-e^{-269T})\},~~~g^{(p)}_{yy}=g^{(p)}_{zz}=
\exp\{6T+\frac{4\chi}{269}(1-e^{-269T})\},
\end{equation} where
superscript $(p)$ denotes $primordial.$ This metric solution shows
at end of primordial inflation the spacetime reaches  to non
vanishing smallest scale in parallel direction with the
cylindrical symmetry axis $x$ but in vertical direction $y=z$ the
inflation continues to next phase such that
\begin{equation}\label{metlim}\lim_{269T>>1}g_{xx}\approx
e^{-\frac{2\chi}{269}},~~~\lim_{269T>>1}g_{yy}=g_{zz}\approx
e^{6T+\frac{4\chi}{269}}.\end{equation} This means that at
duration of the primordial inflation compression  of spacetime is
happened in direction of symmetry axis $x$ while expansion of
space time is happened in its vertical directions $y,z$. In other
words, primordial inflation in this model is similar to  expansion
of cigarette like or pipeline. By substituting (\ref{HNx}) and
definitions $B_1=H_1$ and $\chi=\frac{H_1}{B_c}$ into the
deceleration parameter $q=-1-\frac{\dot{H}}{H^2}$ we obtain
\begin{equation}\label{qNx}q(T)=-1+\frac{269\chi\exp(-269T)}{[
2+\chi\exp(-269T)]^2}.\end{equation} It is easy to check that for
end of primordial inflation the above deceleration parameter
reduces to the following form
\begin{equation}\label{qin}
q_e=\lim_{269T>>1}q(T)\approx -1,
\end{equation} which satisfies the de Sitter epoch
full but for beginning of the primordial inflation we have
\begin{equation}q_0=q(0)=-1+\frac{269\chi}{(
2+\chi)^2},\end{equation} in which the beginning time of the
primordial inflation is chosen to be $t_c=0.$ Negativity condition
on the deceleration parameter at beginning of the primordial
inflation shows the following restriction on the $\chi$ parameter.
\begin{equation}\label{conchi}q_0<0,~~~~-\infty<\chi<0.01509519949.\end{equation}
We assume at end of primordial inflation which is happened after
passing times $T_p>>1,$ the anisotropy can be negligible and the
metric components (\ref{metlim}) reach to the following
approximation
\begin{equation}g_{xx}^{(p)}(T_p)\approx g_{yy}^{(p)}(T_p)\approx e^{N_p},\end{equation}
in which $N_p$ is e-folding parameter of the spacetime at end of
the primordial inflation and so we can write
 \begin{equation}-\frac{2\chi}{269}=N_p=6T_p+4\frac{4\chi}{269}.
\end{equation} This identity gives out \begin{equation} T_p=-\frac{\chi}{269},~~~N_p=2T_p.\end{equation}
By applying these boundary conditions on our general metric solutions given by (\ref{aNx}) and (\ref{bNx}) we now obtain exact
 form of the solutions which satisfy the late time inflation.
\subsection{Late inflationary period}
In this case, the scale of the space time is large and so the
anisotropy can be negligible and so the metric field solutions
(\ref{aNx}) and (\ref{bNx}) will be
\begin{equation}\label{gxxL}g^{(L)}_{xx}=\exp\{2a_c-4b_c+\frac{2\chi}{269}e^{-269T}\},\end{equation}
 and \begin{equation}\label{gyyL}g^{(L)}_{yy}=g^{(L)}_{zz}=\exp\{2a_c+2b_c+6
T-\frac{4\chi}{269}e^{-269T}\},
\end{equation}
where superscript $(L)$ denotes $Late$. If we assume $T_0$ is
beginning time of the late inflationary period in which anisotropy
has more small effects we should use continuity condition of the
metric fields at time $T_0$ as
$g^{(L)}_{xx}(T_0)=g_{yy}^{(L)}(T_0)$ and equality of whose first
derivative
$\frac{dg_{xx}^{(L)}(T_0)}{dT}=\frac{dg_{yy}^{(L)}(T_0)}{dT}$
which by using the above solutions we obtain
 \begin{equation}\label{bc} b_c=-T_0-\frac{1}{269},~~~~\chi=-e^{269T_0}.\end{equation} By substituting (\ref{bc}) into
 the metric field solutions (\ref{gxxL}) and (\ref{gyyL}) we have
\begin{equation}\label{gxxLL}
g^{(L)}_{xx}=\exp\{2a_c+4T_0+\frac{2}{269}(2-e^{-269(T-T_0)})\},
\end{equation} and
\begin{equation}\label{gyyLL}
g^{(L)}_{yy}=g^{(L)}_{zz}=\exp\{2a_c-2T_0+6
T-\frac{2}{269}(1-2e^{-269(T-T_0)})\}.
\end{equation}
By using  the approximation $e^{-269(T-T_0)}\approx1-269(T-T_0)$
we can rewrite the above metric solutions for the times $T>T_0$
such that
\begin{equation}\label{gxyz}g_{xx}^{(L)}\approx g_{yy}^{(L)}=g_{zz}^{(L)}\approx e^{N+2(T-T_0)},
\end{equation} where
\begin{equation}\label{N}
N=2a_c+\frac{2}{269}+4T_0,\end{equation}
  is assumed to be the e-folding parameter of the late time inflation and it should be determined by observational data.  We again remember that
  observations show that the
  current world is old enough such that the e-folding parameter should be a large number between $N\sim\{60\cdots70\}.$ This asymptotic behavior of the
  metric solution (\ref{gxyz}) shows a homogenous and isotropic FRW spacetime in de Sitter epoch in which critical value of the anisotropy velocity $B_c$ behaves as
  an effective cosmological constant such that
\begin{equation}2B_c=\sqrt{\frac{\Lambda}{3}},~~~T=B_ct.\end{equation} However, $N$ is evaluated by observations but how can define  the integral constant
 $a_c$ with other physical parameters of the system? To do so we assume $T_0=T_p$ and $N=N_p$ which means end of primordial inflation is continued with
 beginning of the late time inflation. By regarding this, one can infer
\begin{equation}a_c=-\frac{1}{269}-2T_p.\end{equation}
We end this section by calculating the scalar spectral index
\cite{COL}
\begin{equation}n_s=1+2\eta-6\epsilon,
\end{equation} in which the slow rolling parameters of the inflation $\epsilon$ and $\eta$
mentioned in the previous section take on extended forms with two
variable potential $U(\phi,N_x)$ (two fluid model) as follows
 \begin{equation}\epsilon=\frac{m_p^2}{16\pi}\bigg(\frac{\partial_\phi
   U(\phi,N_x)}{U}+\frac{\partial_{N^x}U(\phi,N_x)}{\phi N_x U
  }\bigg)^2,\end{equation} and \begin{equation}\eta=\frac{m_p^2}{8\pi}\bigg(\frac{\partial_\phi^2 U}{U}+
  2\frac{\partial_{N^x}\partial_\phi U}{\phi N_x U}+\frac{\partial^2_{N^x}U}{\phi^2N_x^2 U}\bigg),\end{equation}
which by substituting (\ref{PotNx}) and chain derivative can be
rewritten as follows
\begin{equation}
\epsilon=\frac{m_p^2}{16\pi}\bigg(\frac{1}{\psi}+\frac{1}{H}-\frac{1}{2B}\bigg)^2\frac{\dot{U}^2}{\phi^2U^2},
\end{equation}
 and
\begin{align}
\eta=&\frac{m^2_p}{8\pi}\bigg\{\bigg(\frac{1}{\phi\psi}+\frac{1}{H}-\frac{1}{2B}\bigg)^2\frac{\ddot{U}}{U}
-\frac{\dot{U}}{U}\bigg[\frac{1}{\phi^3}+\frac{\dot{\psi}}{\phi^2\psi^3}
+\bigg(\frac{2}{\phi\psi}+\frac{1}{H}-\frac{1}{2B}\bigg)\frac{\dot{H}}{H^2}\notag \\
&-\bigg(\frac{2}{\phi\psi}+\frac{1}{2H}-\frac{1}{4B}\bigg)
\frac{\dot{B}}{B^2}\bigg]\bigg\}.
\end{align}
By substituting the critical
 point (\ref{critNx}) and time dependent potential solution (\ref{UNx}) we obtain asymptotic critical values for these slow rolling parameters such that
 \begin{equation}\epsilon_c\approx 0,~~~\eta_c\approx\frac{2}{\pi}\bigg(\frac{30m_p}{\phi_c}\bigg)^2,\end{equation}
 for which the scalar spectral index $n_s$ become
 \begin{equation}n_s\simeq 1+\frac{1}{\pi}\bigg(\frac{60m_p}{\phi_c}\bigg)^2.\end{equation}
In the above relations, we set critical point conditions
$\dot{H}=0=\dot{B}=\dot{U}=\dot{\psi}.$ By regarding the
observation data for the scalar spectral index $n_s\approx 1$, we
infer that the following inequality should be obeyed between
$\phi_c$ and the Planck mass.
\begin{equation}
 \phi_c>>60m_p.
 \end{equation}  In the following section we investigate possible
  stable metric solutions when direction of spatial part of the vector field is in perpendicular to axis of symmetry of the spacetime.
\subsection{Metric solution for $(N_{y}\neq0,N_{t,x,z}=0)$}
   In this case, the equation (\ref{7})
    reduces to the following constraint.\begin{equation}\label{ConNy}\alpha=\frac{i\pi}{2},~~~\beta=\frac{\pi}{2},~~~\gamma=0,~
    ~~N_y=ie^{a+b},\end{equation}
and the field equations $\phi$, $N_{\mu}$ and $G_{\mu\nu}$ reduce to the following formes respectively.
\begin{align}
&2\omega\frac{\ddot{\phi}}{\phi}+2\omega(\dot{a}-2\dot{b})\frac{\dot{\phi}}{\phi}-\omega\frac{\dot{\phi}^2}{\phi^2}-2\dot{a}^2-12
\dot{b}^2-34\dot{a}\dot{b}+8\ddot{a}+\frac{\partial U(\phi,N_y)}{\partial\phi}=0,\label{phiNy}\\
&\frac{\zeta}{\phi}=-54\dot{a}^2-2\dot{b}^2-54\dot{a}\dot{b}-10\ddot{a}-8\ddot{b}-8(\dot{a}+\dot{b})
\bigg(\frac{\dot{\phi}}{\phi}
\bigg)-\frac{\partial U(\phi,N_y)}{\phi N_y\partial
N^y},\label{zetaNy}
\end{align}
and $G_t^t,G_x^x,G_y^y=G_z^z$ will be respectively
\begin{align}
&\ddot{a}+\frac{\omega}{2}\frac{\dot{\phi}^2}{\phi^2}+6\dot{a}^2+3\dot{a}\dot{b}-2\dot{b}^2-\frac{U}{\phi}=0,\label{GttNy}\\
&3\ddot{a}+2\ddot{b}+6\dot{a}^2+9\dot{a}\dot{b}+4\dot{b}^2-\frac{\omega}{2}\frac{\dot{\phi}^2}{\phi^2}-\frac{\ddot{\phi}}{\phi}
-3\dot{a}\frac{\dot{\phi}}{\phi}-\frac{U(\phi,N_y)}{\phi}=0,\label{GxxNy}
\end{align}
   and
   \begin{equation}\label{GyyNy}
   3\ddot{a}-\ddot{b}+6\dot{a}^2+4\dot{b}^2-\frac{\zeta}{\phi}-3\dot{a}\frac{\dot{\phi}}{\phi}
   -\frac{\ddot{\phi}}{\phi}-\frac{\omega}{2}\frac{\dot{\phi}^2}{\phi^2}-\frac{U}{\phi}=0.
\end{equation}
 The above five equations are enough to determine all  quantities $a,b,\phi,\zeta$ and $U(\phi,N_y).$
  By substituting  (\ref{ConNy}) and definitions
  \begin{equation}\label{difNy} \dot{a}=H,~~~\dot{b}=B,~~~\frac{\dot{\phi}}{\phi}=\psi, \end{equation}
  we obtain the following identities for the potential
  \begin{equation}\label{PotNy}\frac{1}{N_y}\frac{\partial U}{\partial N^y}=
  \frac{\partial U}{\partial a}+
  \frac{\partial U}{\partial b}=\bigg(\frac{1}{H}+\frac{1}{B}\bigg)\dot{U},~~~\frac{\partial U}{\partial \phi}=\frac{\dot{U}}{\phi\psi}.
  \end{equation} By using (\ref{difNy}) and (\ref{PotNy}) one can show that the dynamical field equations
  (\ref{phiNy}), (\ref{zetaNy}), (\ref{GttNy}),
   (\ref{GxxNy}) and (\ref{GyyNy}) are transformed to the following first order nonlinear differential equations.
\begin{align}
&\dot{H}=\frac{(-\omega\phi\psi^2+4B^2\phi-6BH\phi-12H^2\phi+2U)}{2\phi},\\
&\dot{B}=-\frac{(9BH\phi+\zeta)}{3\phi},\\
&\dot{\psi}=\frac{(-6\omega\phi\psi^2+30B^2\phi-18BH\phi-36H^2\phi-9H\phi\psi-3\phi\psi^2+6U-2\zeta)}{3\phi},
\end{align}
and
\begin{equation}
\dot{U}=-\frac{HB(-15\omega\phi\psi^2+66B^2\phi+24B\phi\psi-18H^2\phi+24H\phi\psi+30U-5\zeta)}{H+B},
\end{equation} and
\begin{align}
\zeta=&3(4B\omega^2\phi\psi^3+4H\omega^2\phi\psi^3-20B^3\omega\phi\psi-8B^2H\omega\phi\psi+4B^2\omega\phi\psi^2\notag\\
&+36BH^2\omega\phi\psi+3BH\omega\phi\psi^2+5B\omega\phi\psi^3+24H^3\omega\phi\psi+4H^2\omega\phi\psi^2+5H\omega\phi\psi^3\notag\\
&+22B^3H\phi-4B^3\phi\psi+62B^2H\phi\psi-6BH^3\phi+116BH^2\phi\psi+50H^3\phi\psi-4BU\omega\psi\notag \\
&-4HU\omega\psi+10BHU-8BU\psi-8HU\psi)/(-4B\omega\psi-4H\omega\psi+5BH).
\end{align}
Similar to previous
sections we solve the equations
$\dot{H}=0=\dot{B}=\dot{\psi}=\dot{U}$ to obtain five critical
manifolds which they are collected in the following table.
\begin{center}
\begin{tabular}{|c|c|c|c|c|}
  \hline
  $\omega$ & x & y & p & q \\
  \hline
  -2.38189179& -0.81852617 & -0.92679835& 7.36673550 & -1.45863725 \\
  0.82291508& -0.50652226 & 1.75767870 & 4.55870034 & -0.70900689 \\
  2.41213514 & -0.71882277 & 1.08689952 & 6.46940495 & 0.36855696 \\
  2.53481716 & -0.06704170& -1.54604186 & 0.60337534 & -186521838 \\
  0.50000000 & -0.42183288 & 2.00000000 & 3.79649596 & -1.19784076 \\
  \hline
\end{tabular}\end{center}
 where we defined
\begin{equation}H_c=xB_c,~~~\psi_c=yB_c,~~~\zeta_c=p\phi_cB_c^2,~~~~U_c=q\phi_cB_c^2.\end{equation}
Physical boundary conditions for an accelerating expanding
universe let us to choose $H_c>0$ for which time trajectory of the
Brans Dicke scalar field should be raising function versus the
cosmic times and so we should choose $\psi_c>0.$ These conditions
lead to say that physical critical manifolds given in the above
table are just $(x,y)>0$ for $B_c>0$ or $(x,y)<0$ for $B_c<0.$ By
regarding the latter conditions and by looking at the numerical
values of the table, we infer that
$\omega=\{-2.381891793,2.534817164\}$ are physical choices for
which both of the dimensionless critical manifolds $x$ and $y$
have negative numeric values synchronously and with $B_c<0$ they
satisfy the physical boundary conditions $(H_c,\psi_c)>0.$  To
determine whether this critical point  give stable metric
solutions, we need to determine sign of the eigenvalues. This is
done by solving the secular equation of the Jacobi matrix for the
above dynamical field equations similar to the previous sections.
By doing this, we obtain for eigenvalues
\begin{align}
&\frac{\sigma_{1}^1}{B_c}=0.1512854848,&&\frac{\sigma_{2}^1}{B_c}=40.08736104,\label{z1c}\\
&\frac{\sigma_{3}^1}{B_c}=4.811325390+1.938390681i,&&
\frac{\sigma_{4}^1}{B_c}=4.811325390-1.938390681i, \notag
\end{align}
for $\omega=-2.381891793$ and
\begin{align}
&\frac{\sigma_{1}^1}{B_c}=0.2146088815,&&\frac{\sigma_{2}^1}{B_c}=11.13107522,\label{z2c}\\
&\frac{\sigma_{3}^1}{B_c}=46294.82001,&&\frac{\sigma_{4}^1}{B_c}=-46288.47253,\notag
\end{align}
for $\omega=2.534817164.$ It is easy to see that the critical
point (\ref{z1c}) shows spiral source (unstable) and (\ref{z2c})
shows quasi stable nature for the metric solutions of this
subsection because for the former case, real part of all
eigenvalues have positive numeric values while for the latter case
 at least one of the eigenvalues has negative real value. Hence, we end this section by plotting arrow diagram just for the latter quasi
 stable solution in Figure 7.
\section{Concluding remark}
In this work, we used the modified Brans Dicke scalar vector
tensor gravity to study anisotropic Bianchi I cosmology. We solved
dynamical field equations for different directions of the timelike
dynamical vector field which interacts nonminimally with the Brans
Dicke scalar field and the metric field. We succeeded to obtain
analytic metric field solutions for epoch of primordial and late
time inflations. In this model, in phase of primordial inflation
the space time expansion is similar to pipeline geometry while in
the late time inflation expansion is isotropic de Sitter epoch
asymptotically. Time trajectories for anisotropy part of the
metric field and the Hubble parameter and the Brans Dicke scalar
field are not negligible in the primordial inflation but are at
the late inflationary period. In the late inflationary period,
critical value of velocity of the anisotropy behaves as an
effective cosmological constant which can be claimed that it is
really origin of the unknown cosmological parameter. By applying
dynamical system approach, we investigated stability conditions of
the solutions and obtained stable (un stable) nature in phase
space when spatial directions of the timelike vector field is
parallel (perpendicular) to the axes of symmetry of the spacetime.
Negativity sign of the barotropic index which can be considered
apparently as characterization of dark matter/energy, is in fact
depended to direction of the vector field in the 4 dimensional
Bianchi I line element. It has positive (negative) numeric value
when spatial components of the vector field is ( is not)
eliminated. In fact, when spatial components of the vector field
is parallel to the symmetry axis of the spacetime our metric
solutions have two different branches. One branch describes
accelerating expansion of a Bianchi I cosmology with a stable
nature in phase space while the second branch which describes a
collapsing metric solution with quasi stable nature can be
considered as reheating phase of the expansion. As extension of
this work, we like to study dynamics of reheating phase with more
details of this SVT Brans Dicke Bianchi I cosmology in our next
work.

\begin{figure}[htp] \centering
\subfigure[{}]{\label{aa}
\includegraphics[width=.45\textwidth]{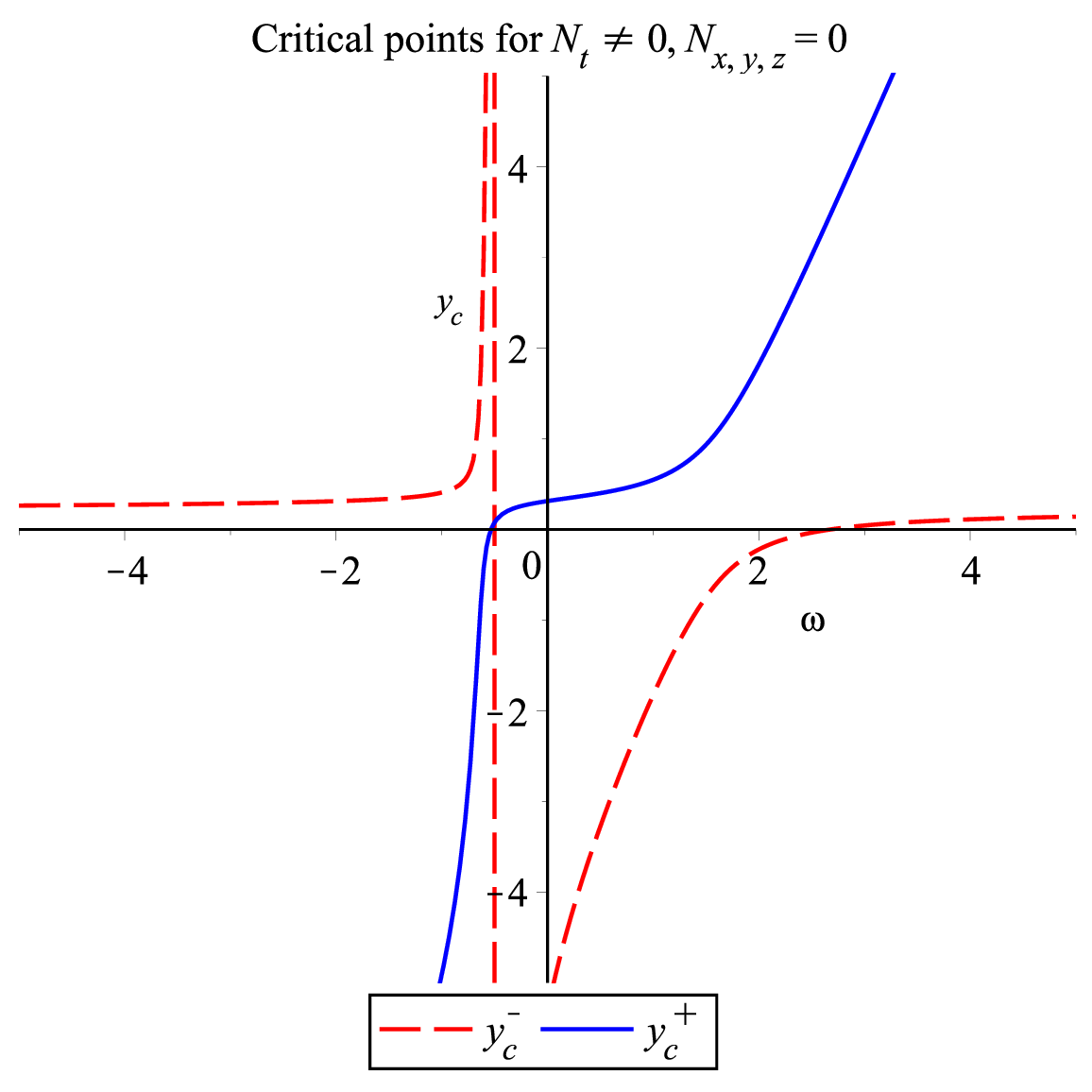}}
\subfigure[{}]{\label{cc}
\includegraphics[width=.45\textwidth]{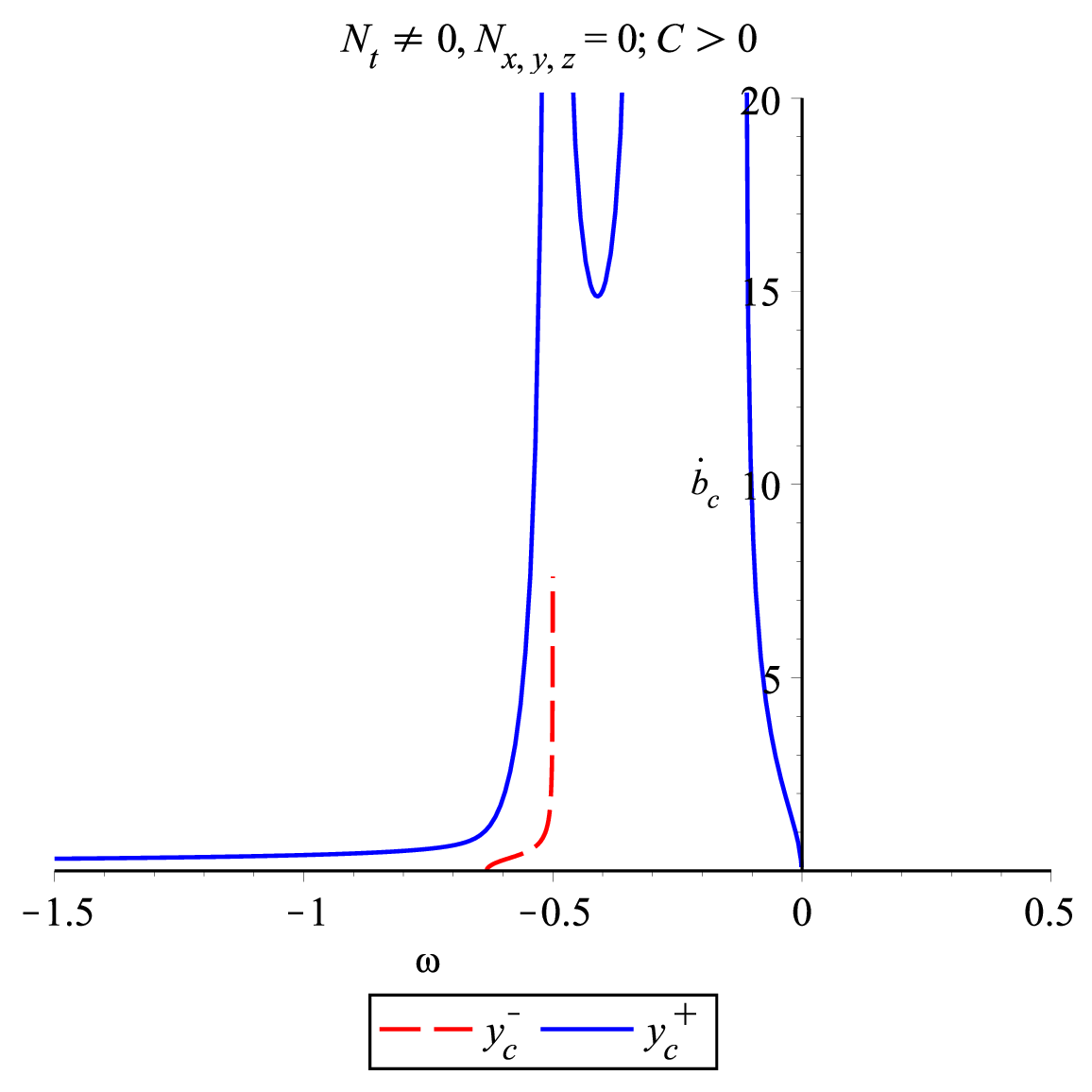}}
\subfigure[{}]{\label{dd}
\includegraphics[width=.45\textwidth]{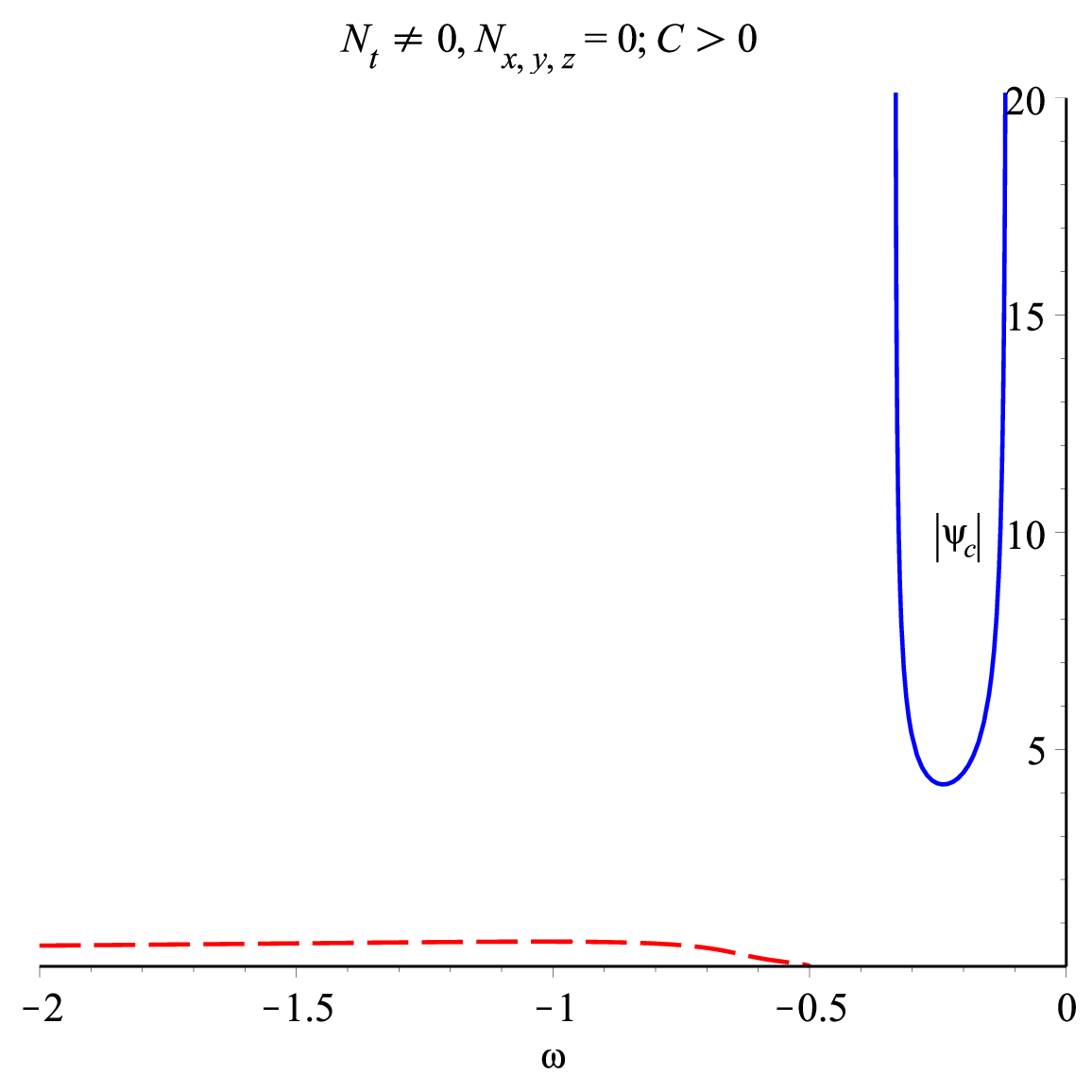}}
\subfigure[{}]{\label{cc}
\includegraphics[width=.45\textwidth]{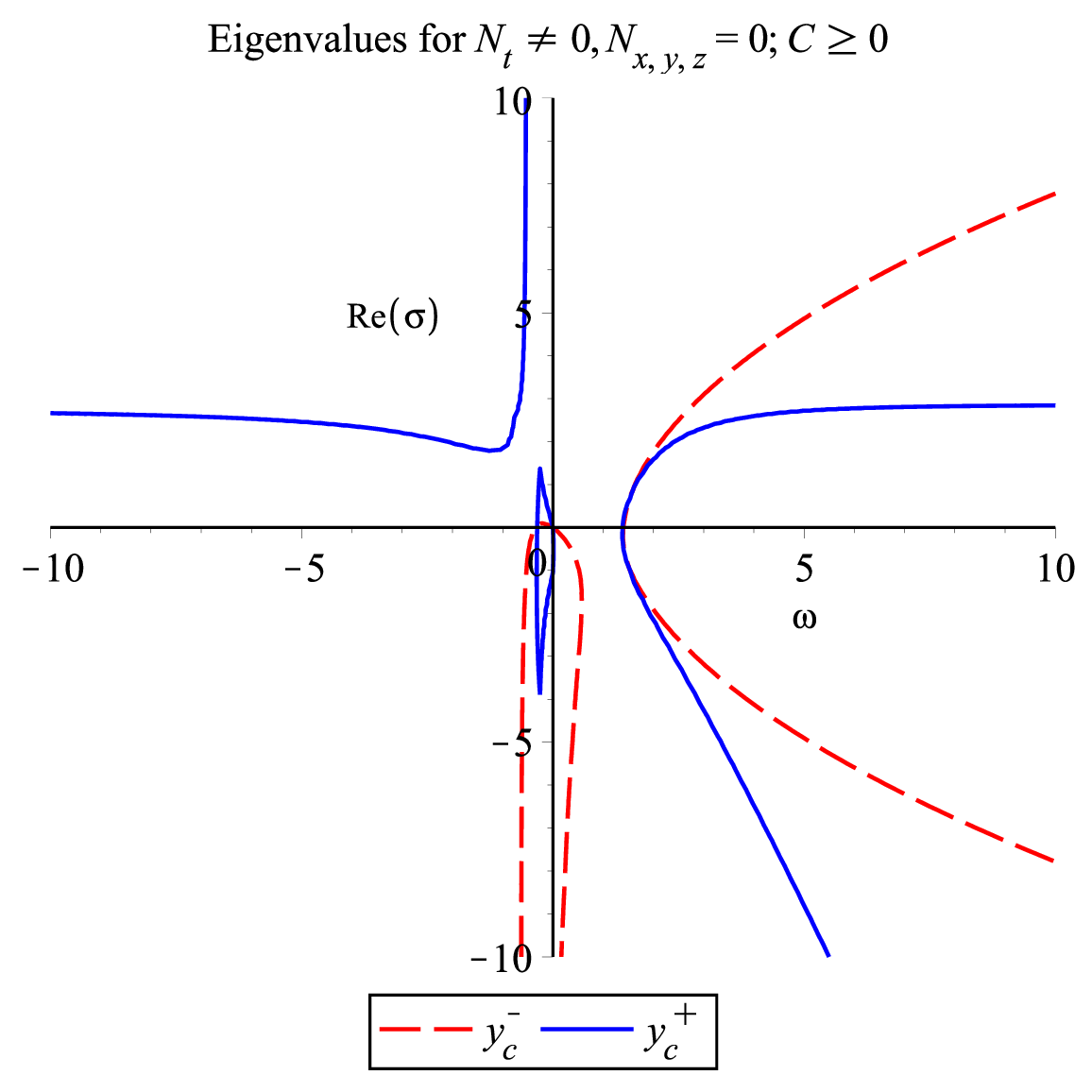}}
 \caption{(a) Numeric values of the critical manifolds vs $\omega$ for
$N_t=1,N_{x,y,z}=0,$ (b) Critical time trajectory of the
anisotropy vs $\omega$ with repeller potential $C>0$ for $N_t=1,
N_{x,y,z}=0,$ (c) Critical time trajectory of the Brans Dicke
scalar field vs $\omega$ with repeller potential $C>0$ for
$N_t=1,N_{x,y,z}=0$ and (d) Real part of eigenvalues vs $\omega$
for $N_t=1,N_{x,y,z}=0$ and $C\geq0$} \label{l}
\end{figure}

\begin{figure}[htp] \centering
\hspace{3mm} \subfigure[{}]{\label{bb}
\includegraphics[width=.45\textwidth]{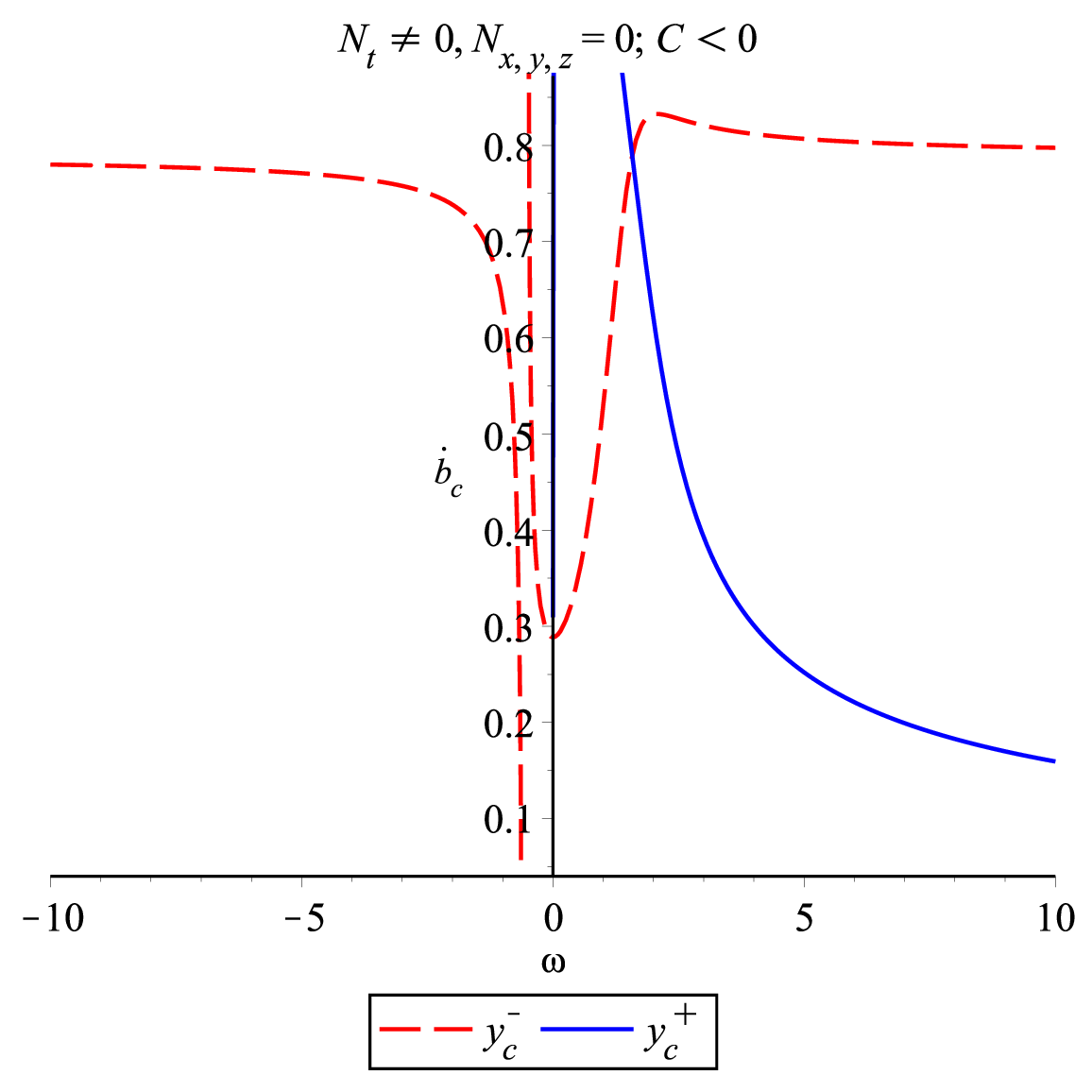}}
\hspace{3mm} \subfigure[{}]{\label{dd}
\includegraphics[width=.45\textwidth]{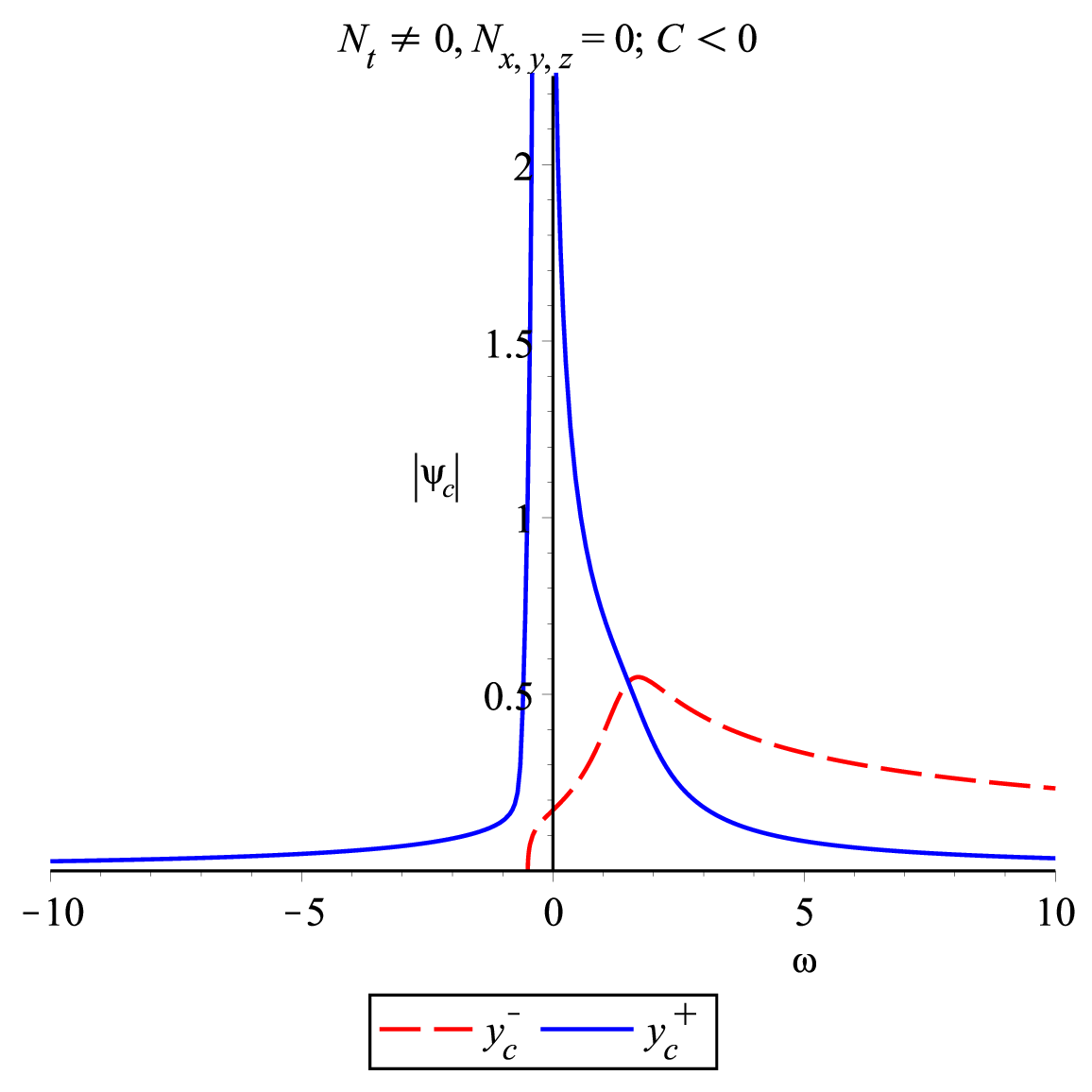}}
\hspace{3mm} \subfigure[{}]{\label{dd}
\includegraphics[width=.45\textwidth]{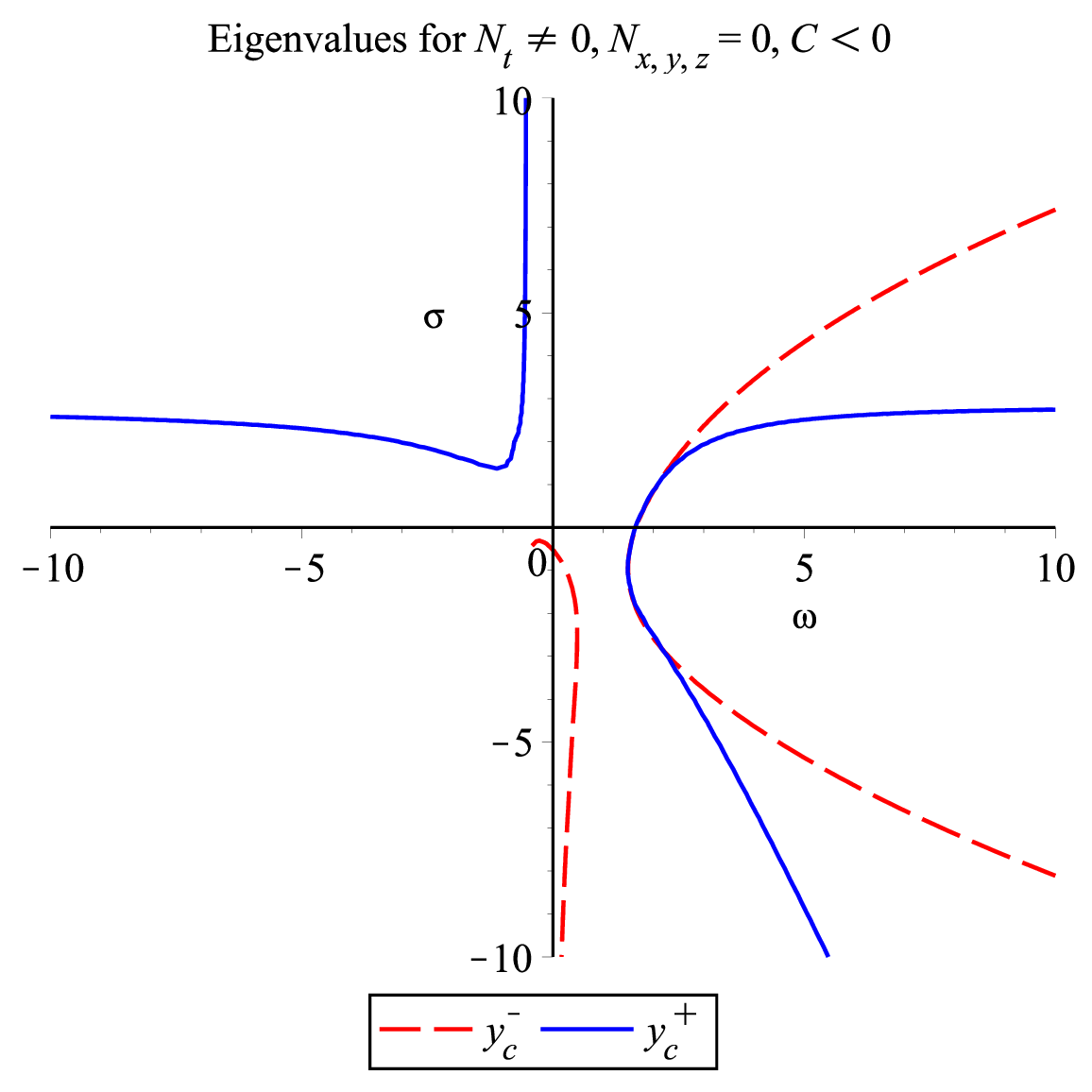}}
 \caption{(a) Critical time trajectory of the
anisotropy vs $\omega$ with absorber potential $C<0$ for $N_t=1,
N_{x,y,z}=0,$  (b) Critical time trajectory of the Brans Dicke
scalar field vs $\omega<0$ with absorber potential $C<0$ for
$N_t=1,N_{x,y,z}=0$ (c) Eigenvalues for $N_t=1,N_{x,y,z}=0$ with
attractor potential $C<0$} \label{l}
\end{figure}

\begin{figure}[htp] \centering
\hspace{3mm} \subfigure[{}]{\label{aa}
\includegraphics[width=.45\textwidth]{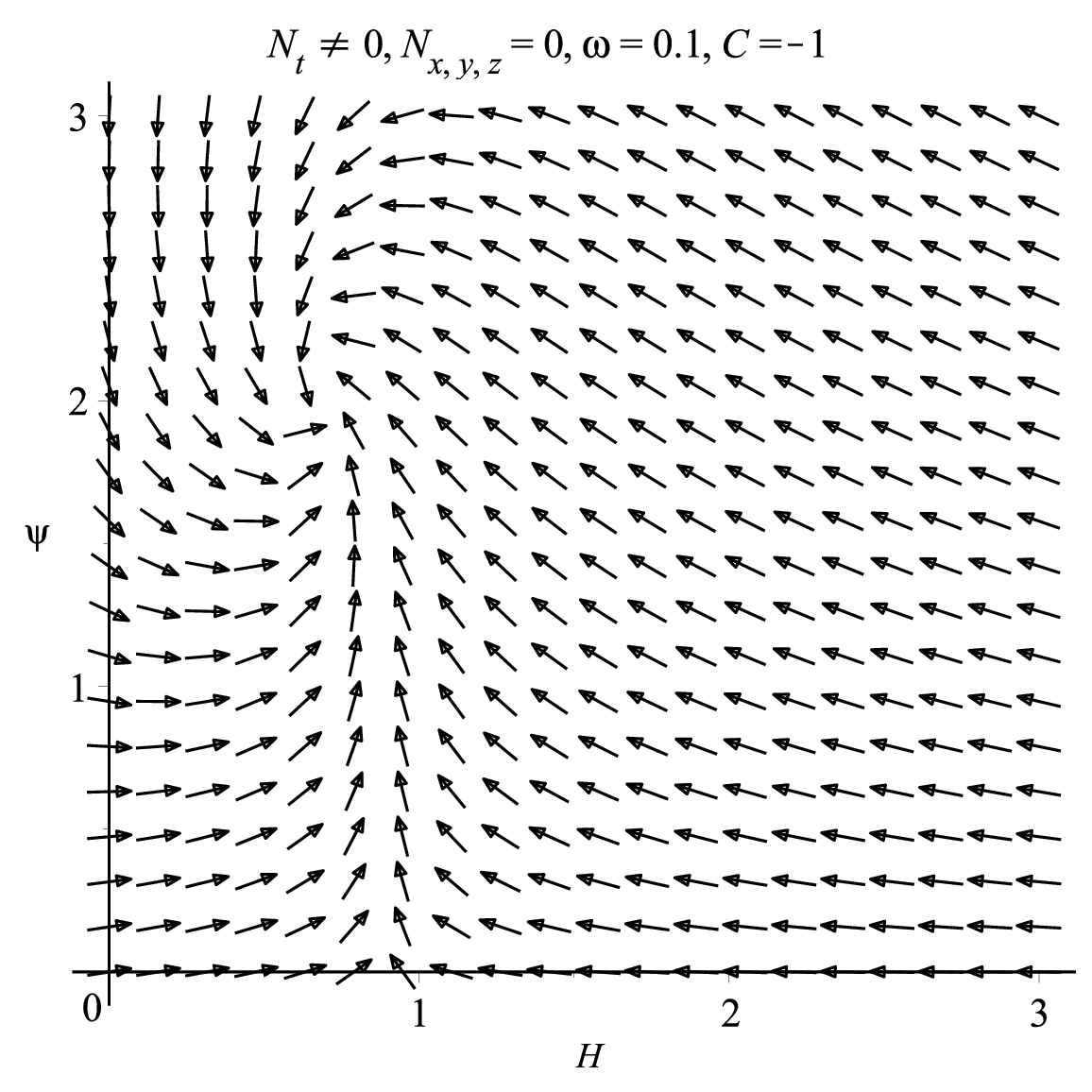}}
\hspace{3mm} \subfigure[{}]{\label{aa}
\includegraphics[width=.45\textwidth]{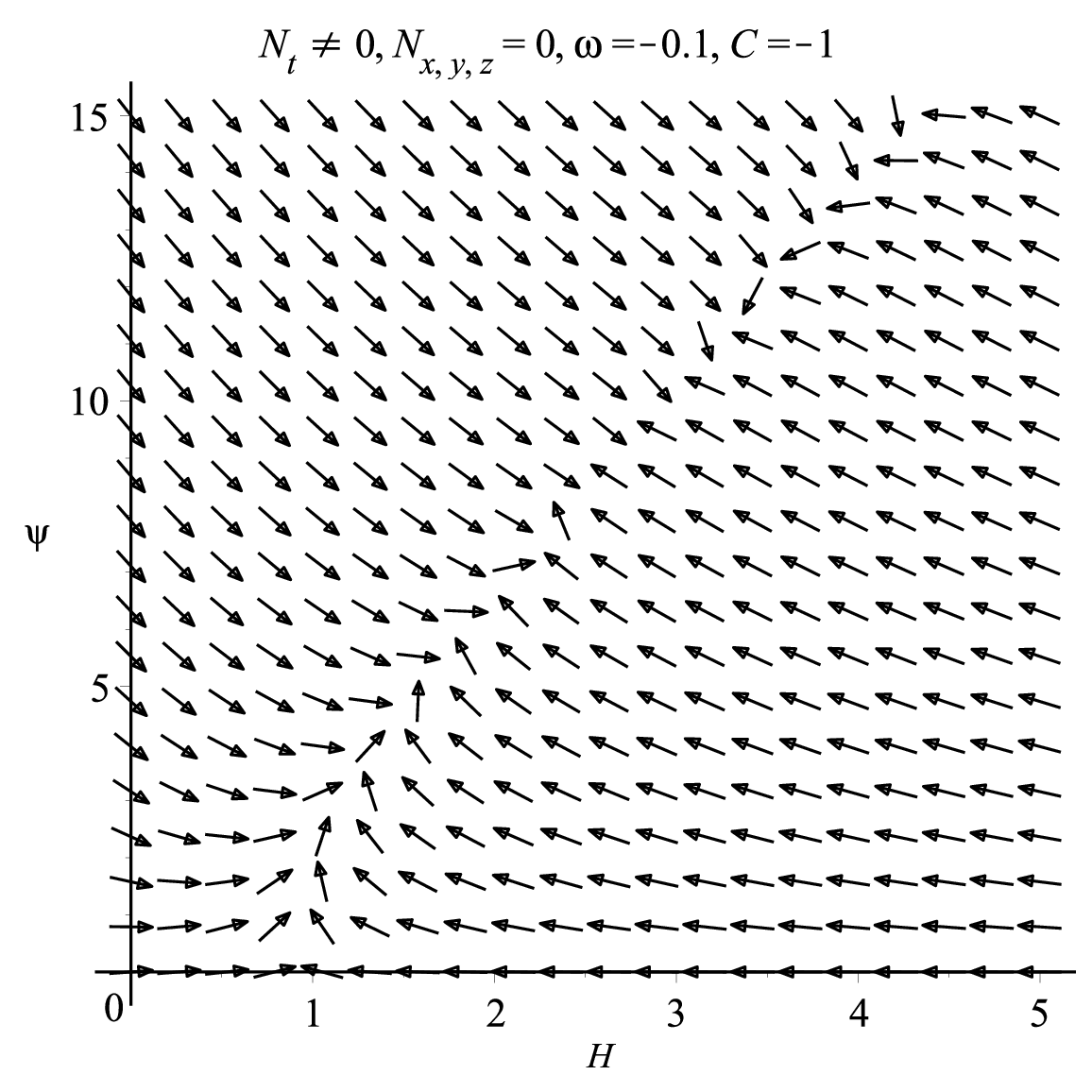}}
\hspace{3mm} \subfigure[{}]{\label{aa}
\includegraphics[width=.45\textwidth]{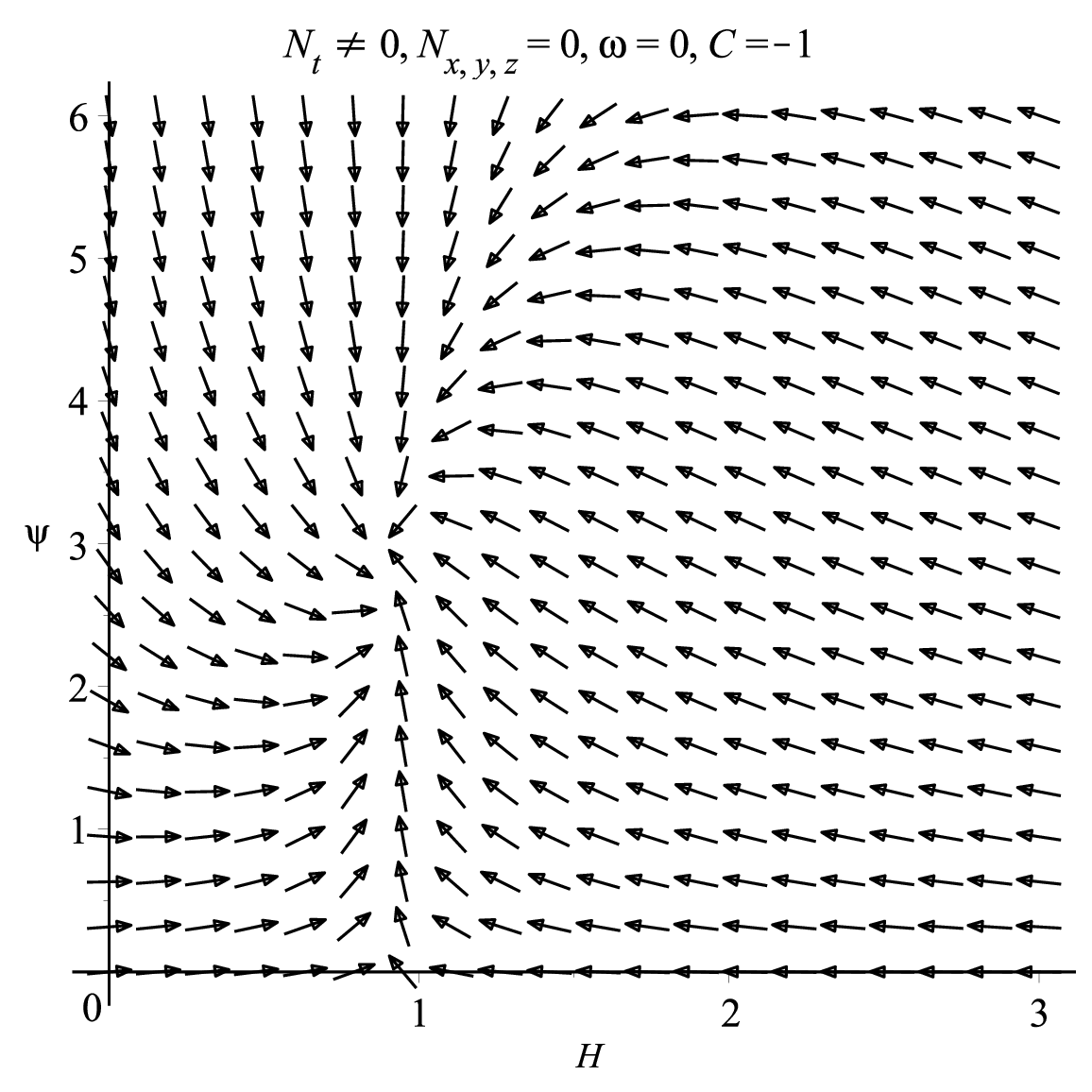}}
 \caption{Arrow diagrams for attractor potential $C=-1$ for $\omega=0.1$ (a), $\omega=-0.1$ (b) and $\omega=0$
  (c) in case $N_t\neq0,N_{x,y,z}=0$} \label{l}
\end{figure}

\begin{figure}[htp] \centering
\hspace{3mm} \subfigure[{}]{\label{cc}
\includegraphics[width=.45\textwidth]{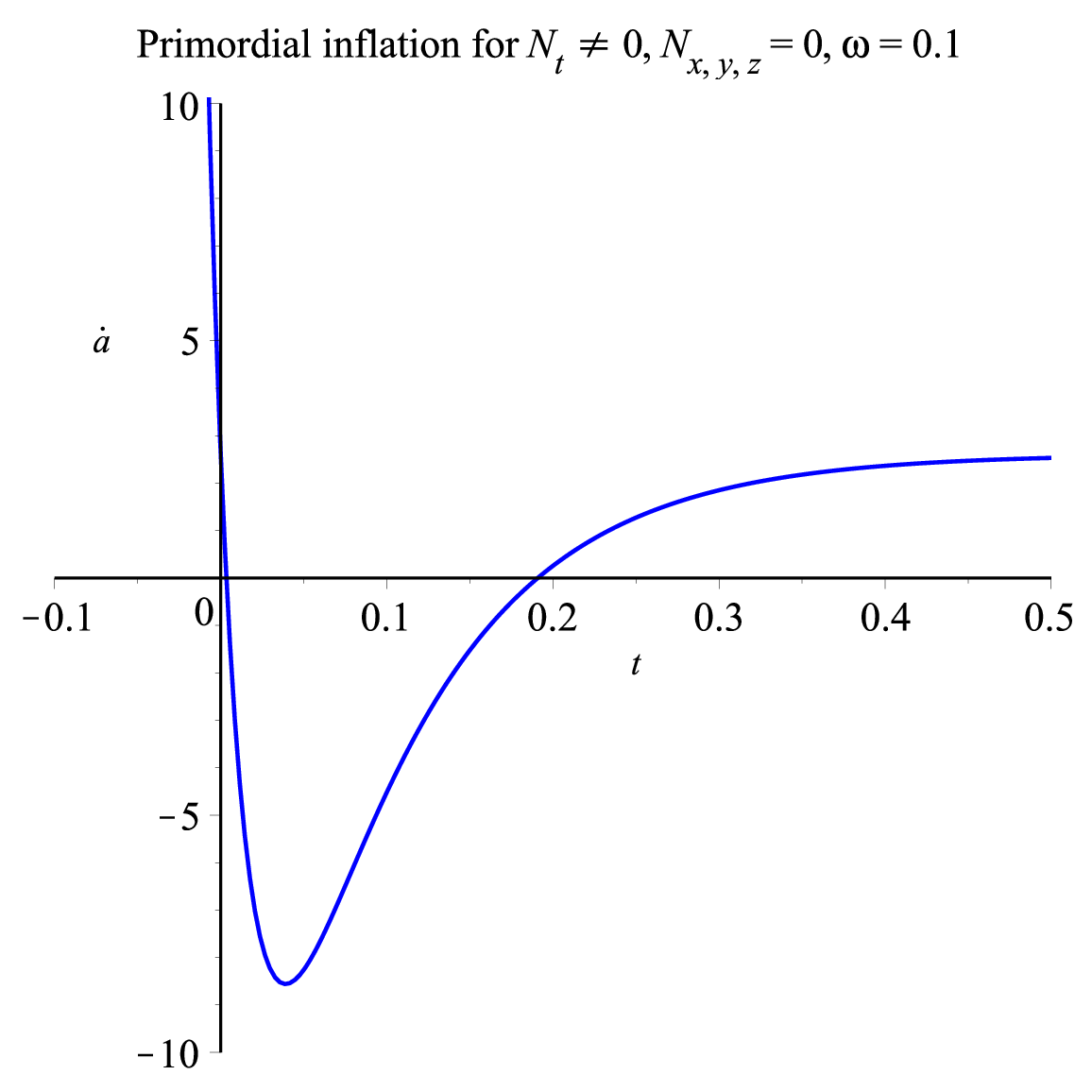}}
\hspace{3mm} \subfigure[{}]{\label{dd}
\includegraphics[width=.45\textwidth]{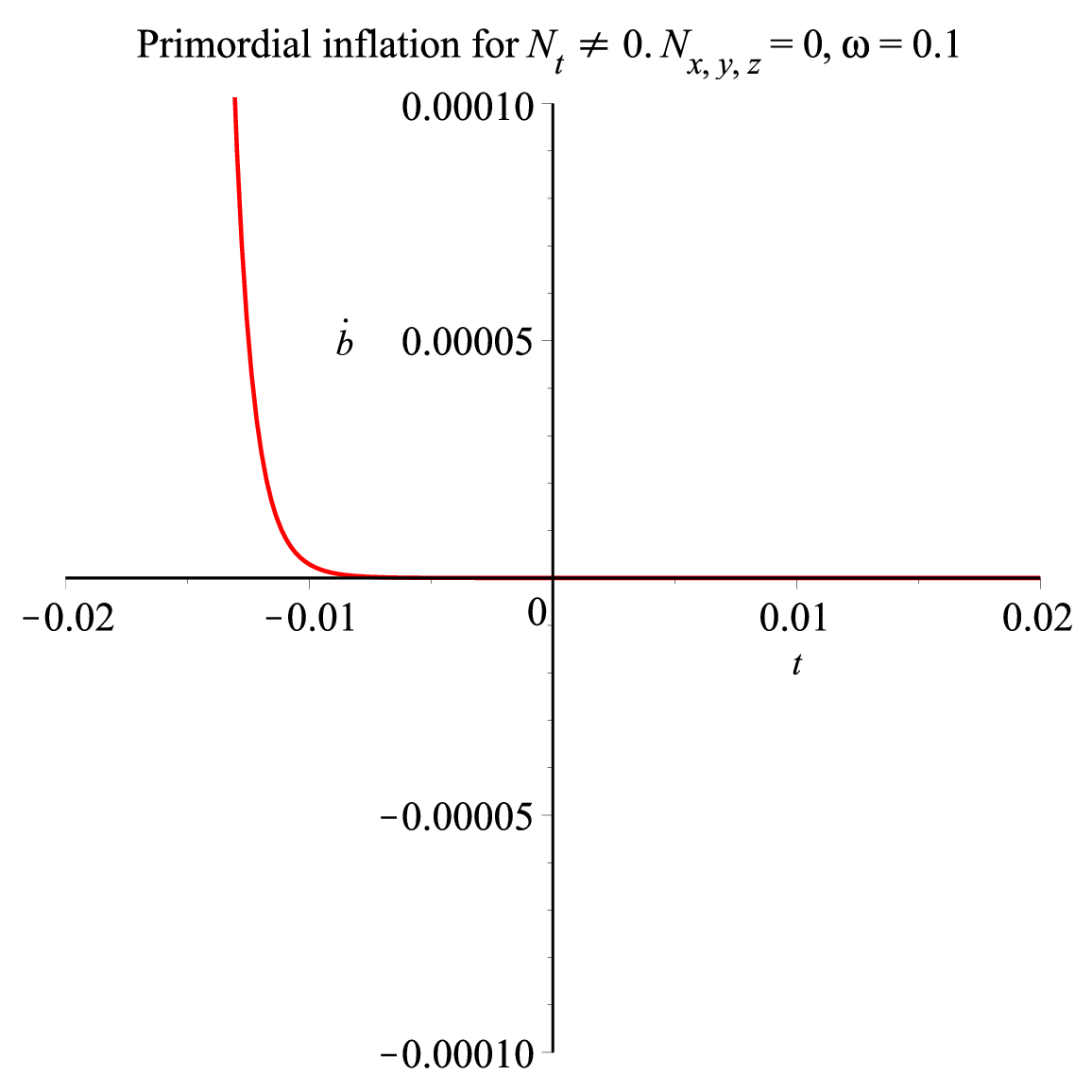}}
\hspace{3mm} \subfigure[{}]{\label{cc}
\includegraphics[width=.45\textwidth]{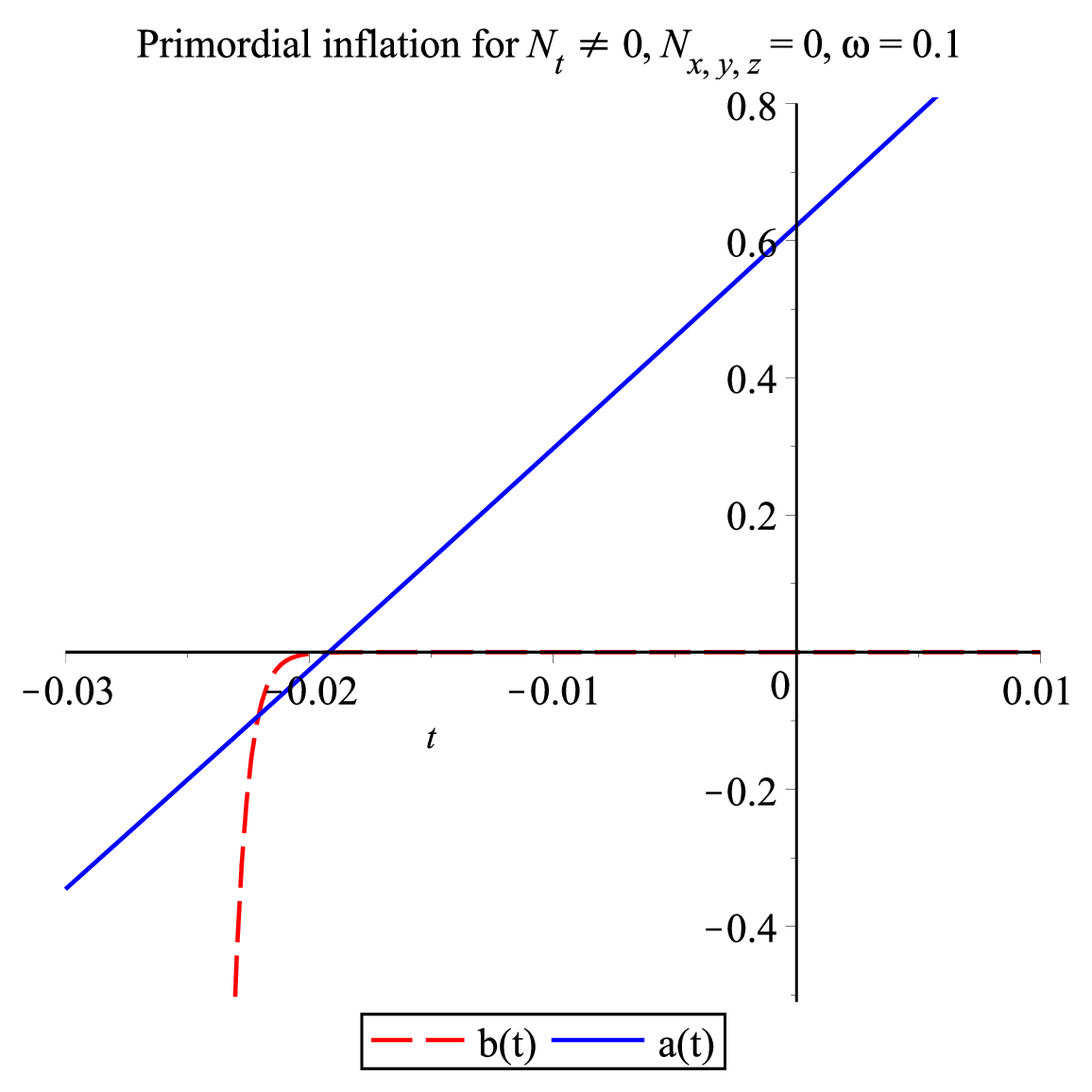}}
\hspace{3mm} \subfigure[{}]{\label{dd}
\includegraphics[width=.45\textwidth]{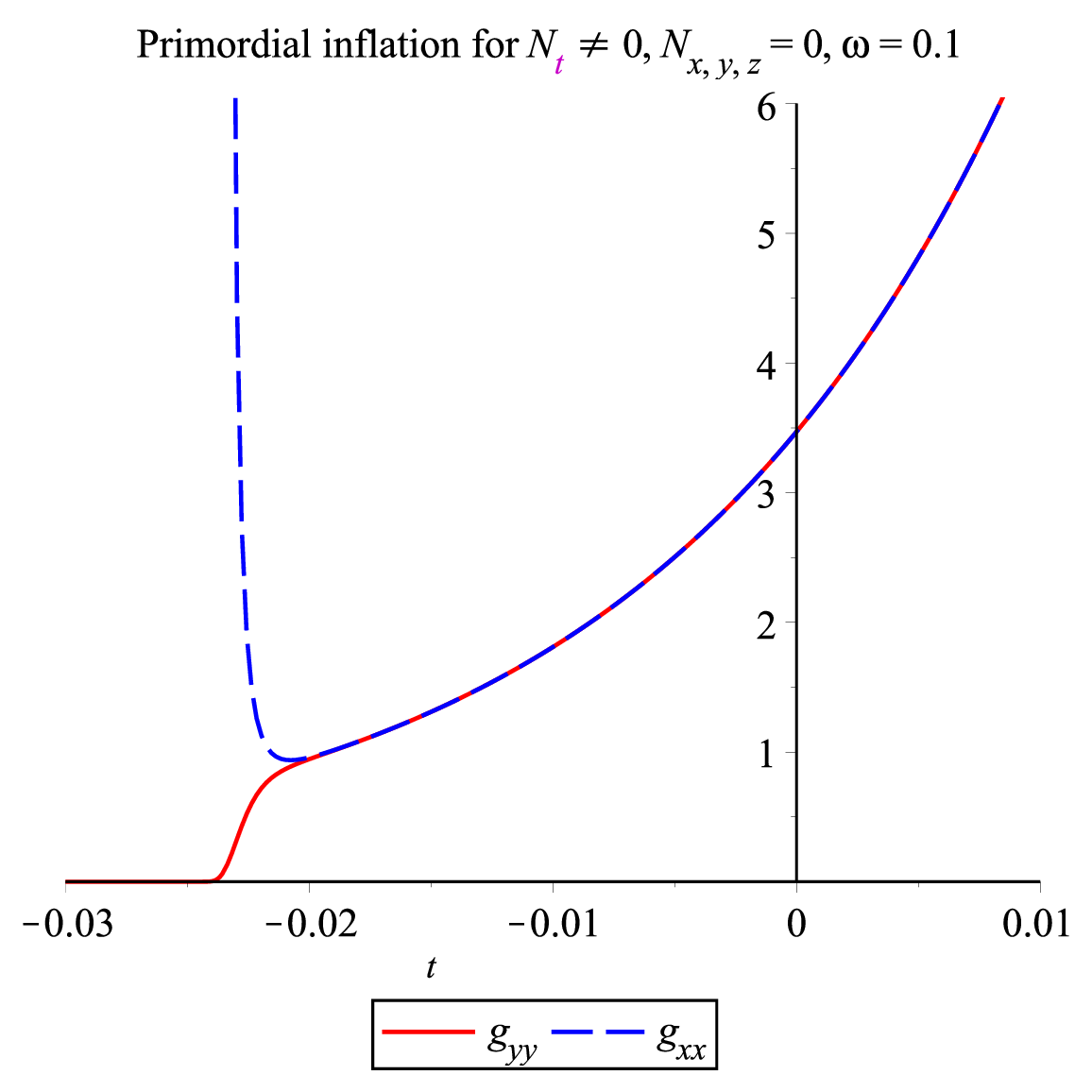}}
\caption{ (a,b)Isotropic and anisotropic time trajectories and (c)
isotropic and anisotropic scale factors and (d) directional metric
components vs  cosmic time $t$ for primordial inflation in case
$N_t=1,N_{x,y,z}=0$ for stable solution $\omega=0.1$ given in the
table 1} \label{l}
\end{figure}

\begin{figure}[htp] \centering
\hspace{3mm} \subfigure[{}]{\label{cc}
\includegraphics[width=.45\textwidth]{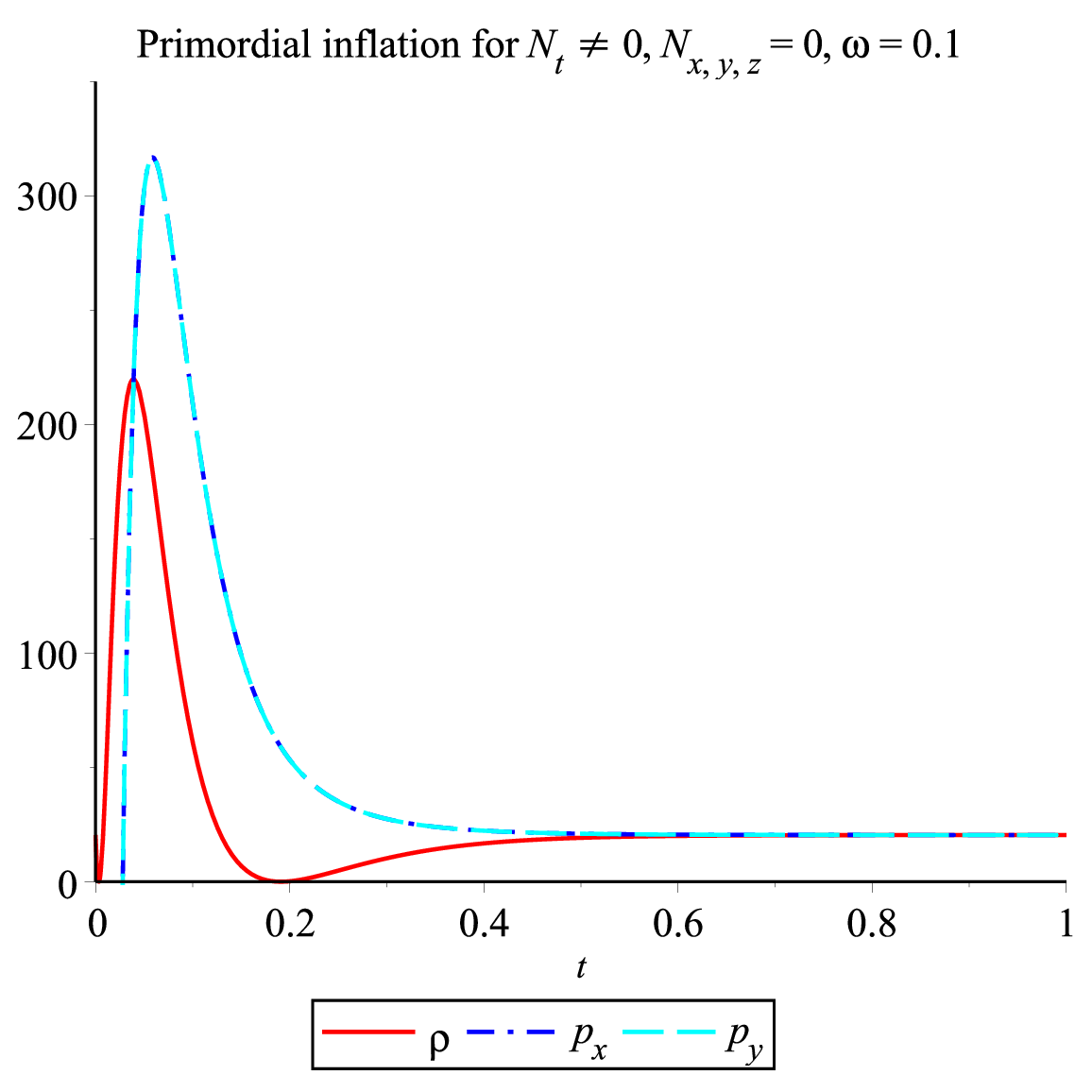}}
\hspace{3mm} \subfigure[{}]{\label{dd}
\includegraphics[width=.45\textwidth]{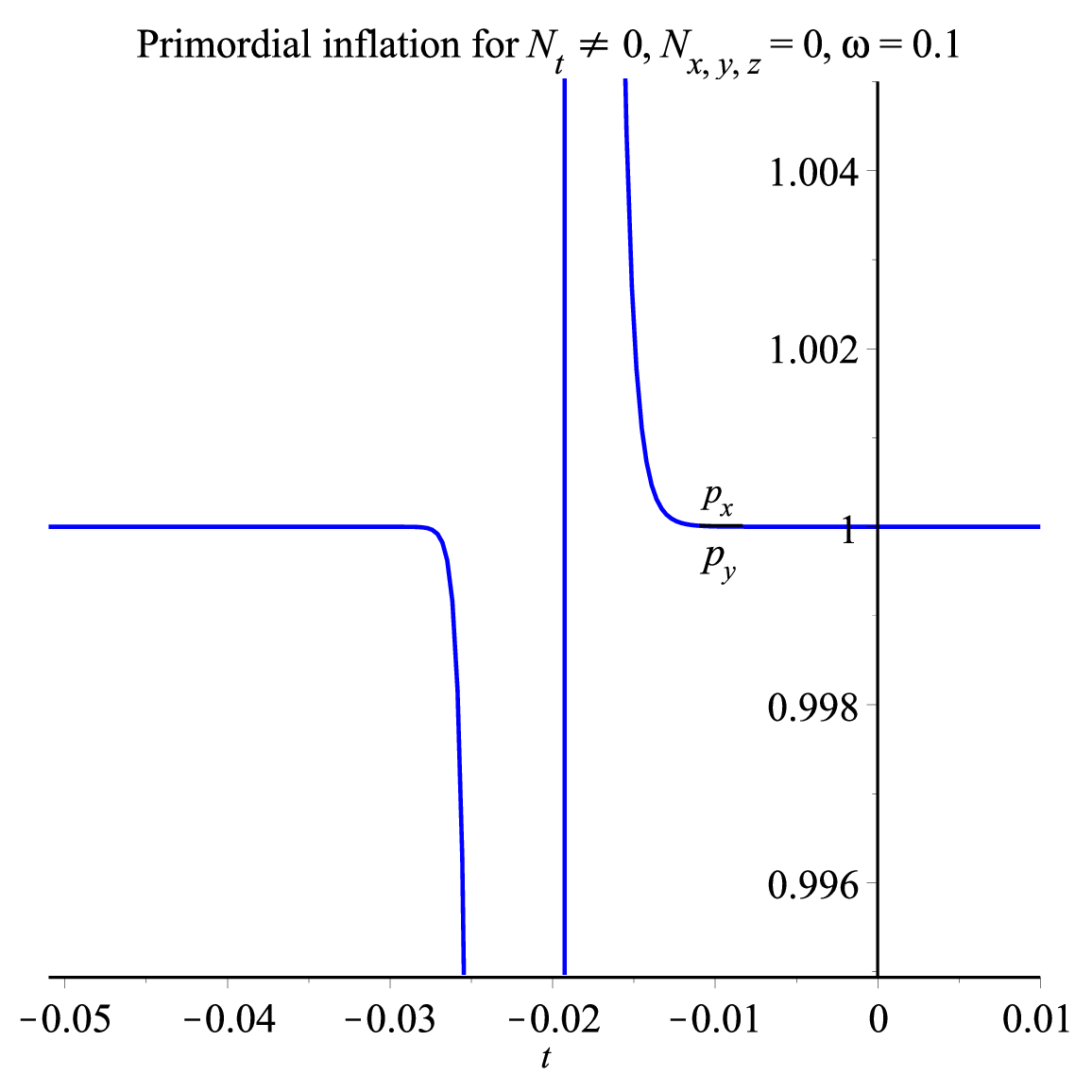}}
\hspace{3mm} \subfigure[{}]{\label{cc}
\includegraphics[width=.45\textwidth]{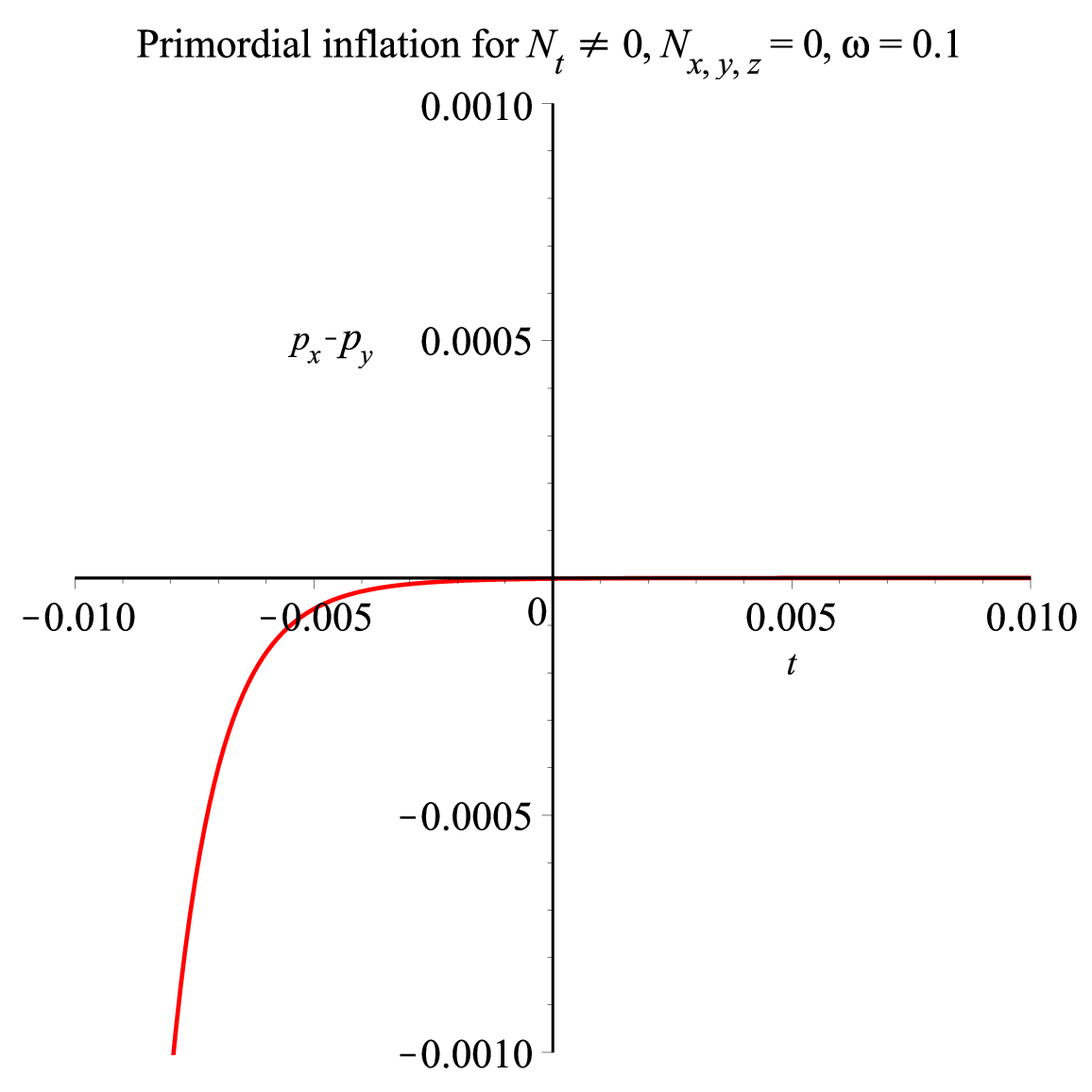}}
\hspace{3mm} \subfigure[{}]{\label{dd}
\includegraphics[width=.45\textwidth]{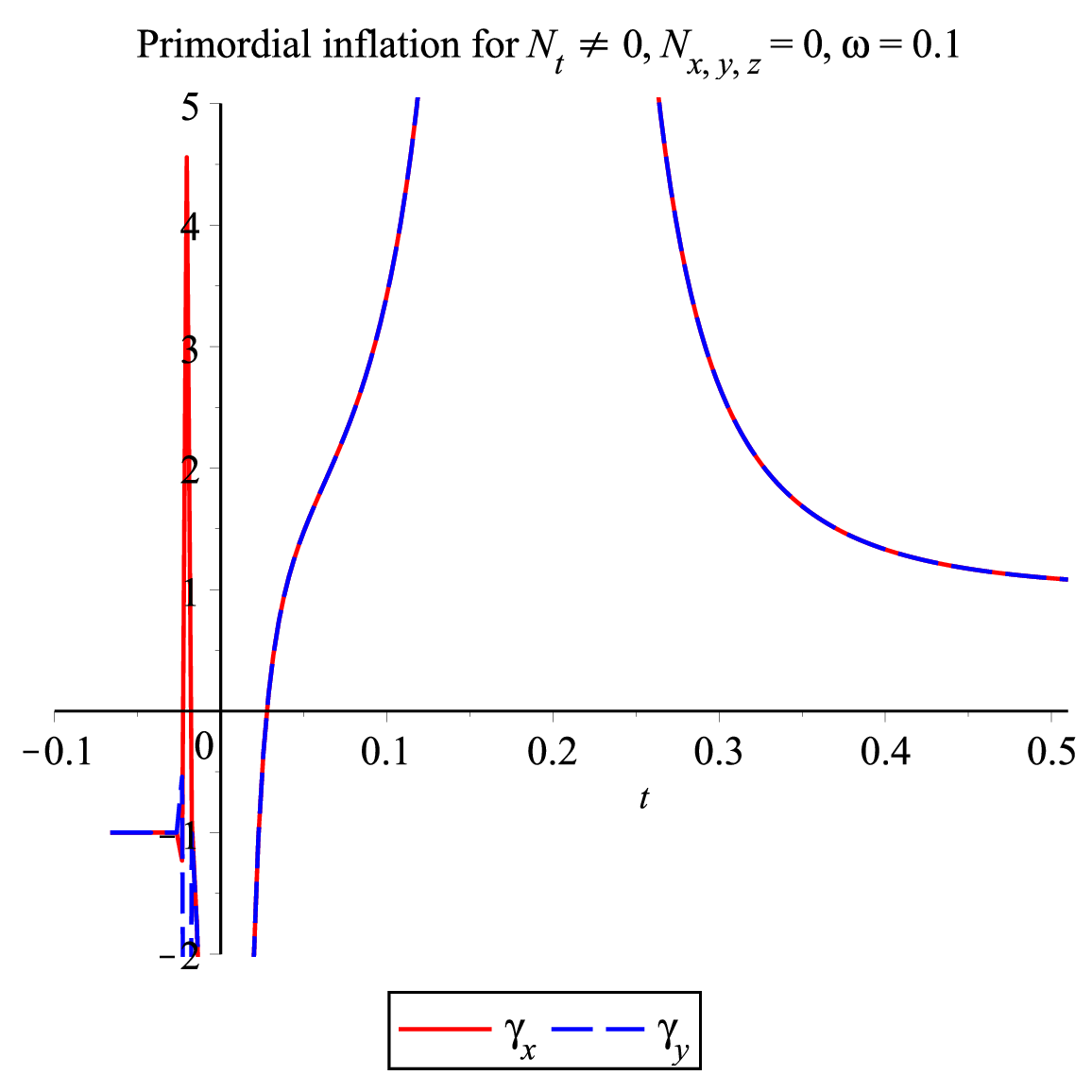}}
\caption{ (a) Re-scaled density and directional pressures vs the
dimensionless cosmic time $t$ and (b) directional barotropic
indexes vs the dimensionless cosmic time $t$ for primordial
inflation in case $N_t=1,N_{x,y,z}=0$ } \label{l}
\end{figure}

\begin{figure}[htp] \centering
\hspace{3mm} \subfigure[{}]{\label{aa}
\includegraphics[width=.45\textwidth]{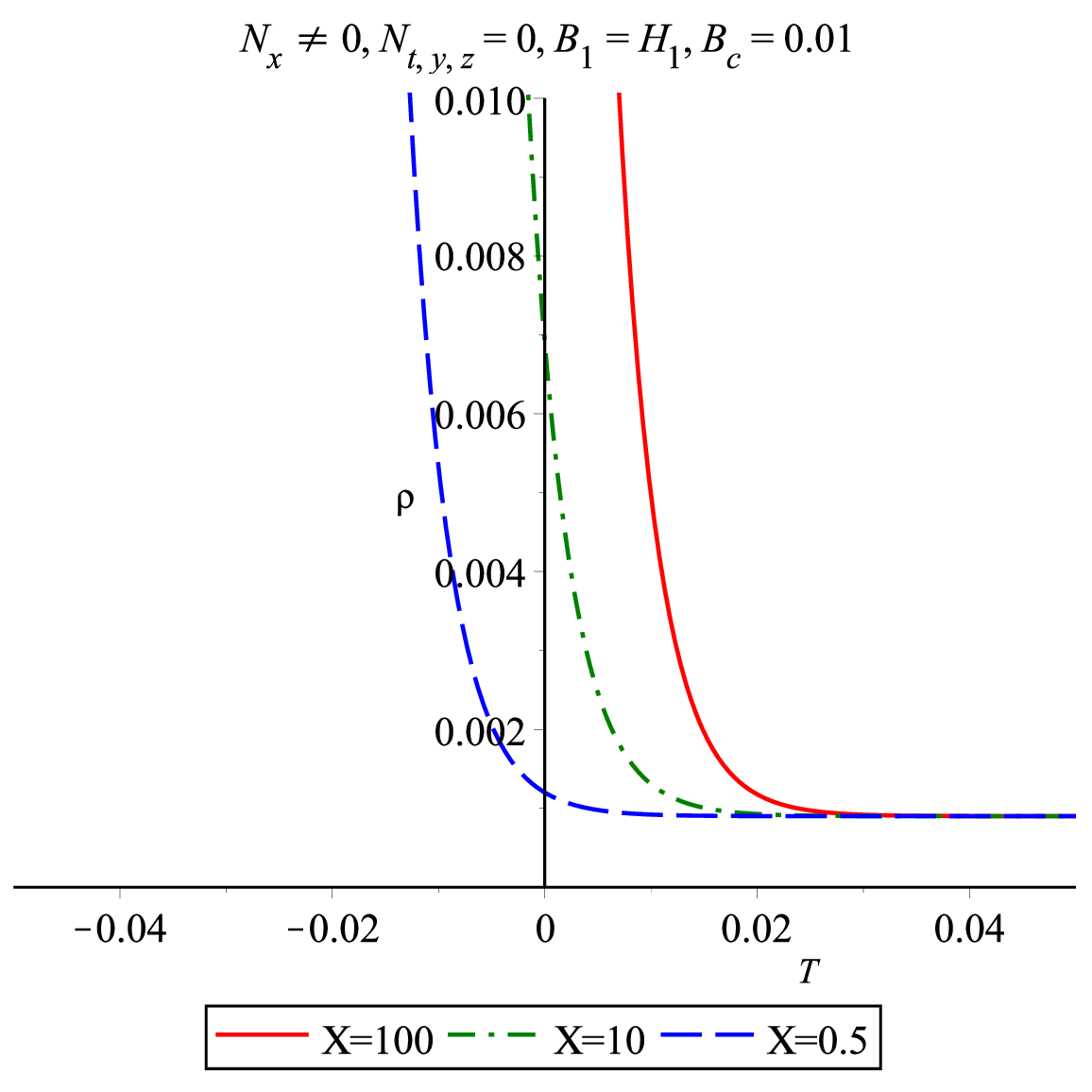}}
\hspace{3mm} \subfigure[{}]{\label{bb}
\includegraphics[width=.45\textwidth]{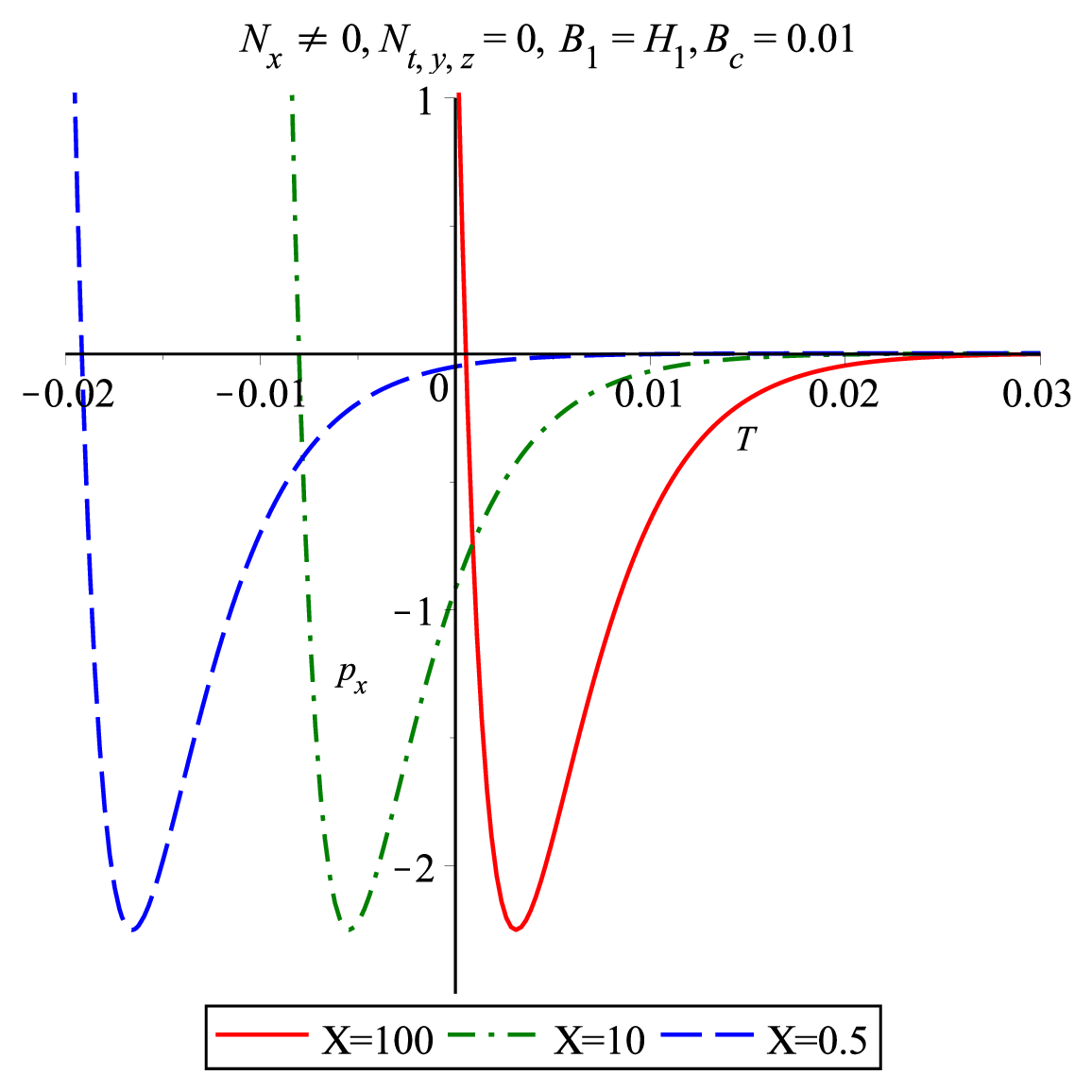}}
\hspace{3mm} \subfigure[{}]{\label{cc}
\includegraphics[width=.45\textwidth]{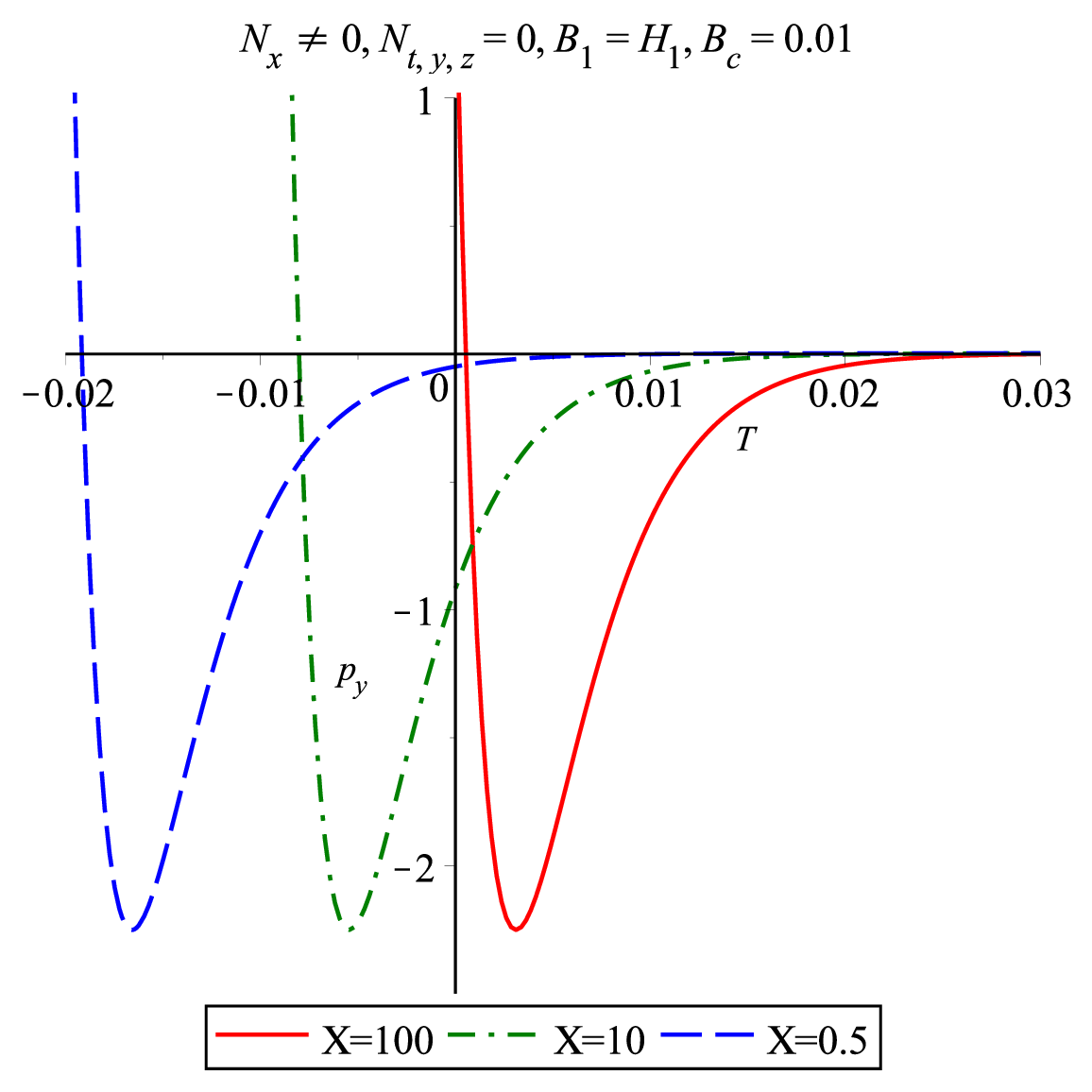}}
\hspace{3mm} \subfigure[{}]{\label{dd}
\includegraphics[width=.45\textwidth]{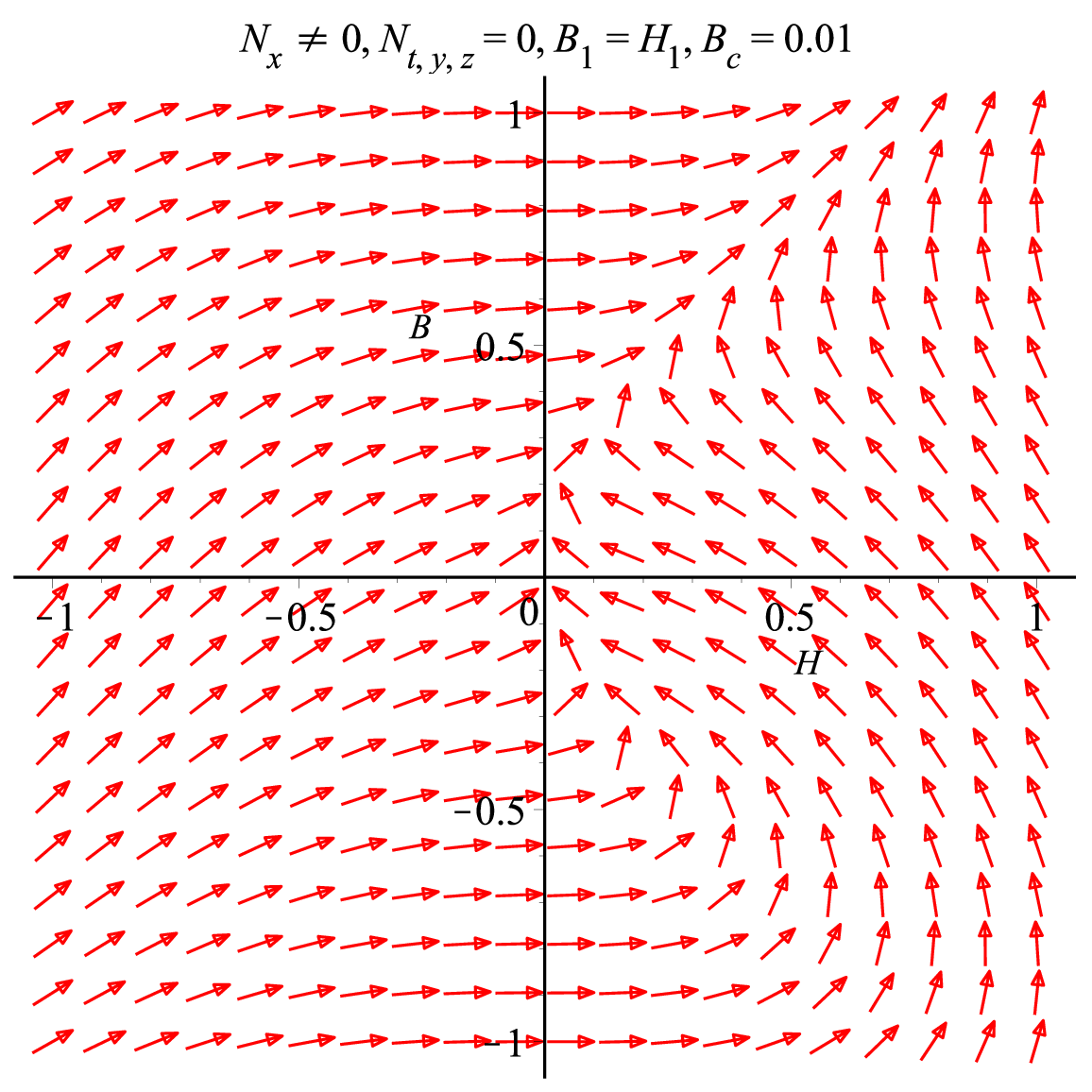}}
 \caption{(a) Diagram of the density vs dimensionless cosmic time $T$ for $N_{x}\neq0,~N_{t,y,z}=0,$ (b) pressure in $x$ direction is plotted vs $T$
 for $N_x\neq0,N_{t,y,z}=0,$ (c) Diagram of $y$ direction pressure is plotted vs $T$ for $N_x\neq0,N_{t,y,z}=0$ and (d) Arrow diagrams for case
  $N_x\neq0,N_{t,y,z}=0$ with sink hole nature.} \label{l}
\end{figure}

\begin{figure}[htp] \centering
\hspace{3mm} \subfigure[{}]{\label{aa}
\includegraphics[width=.45\textwidth]{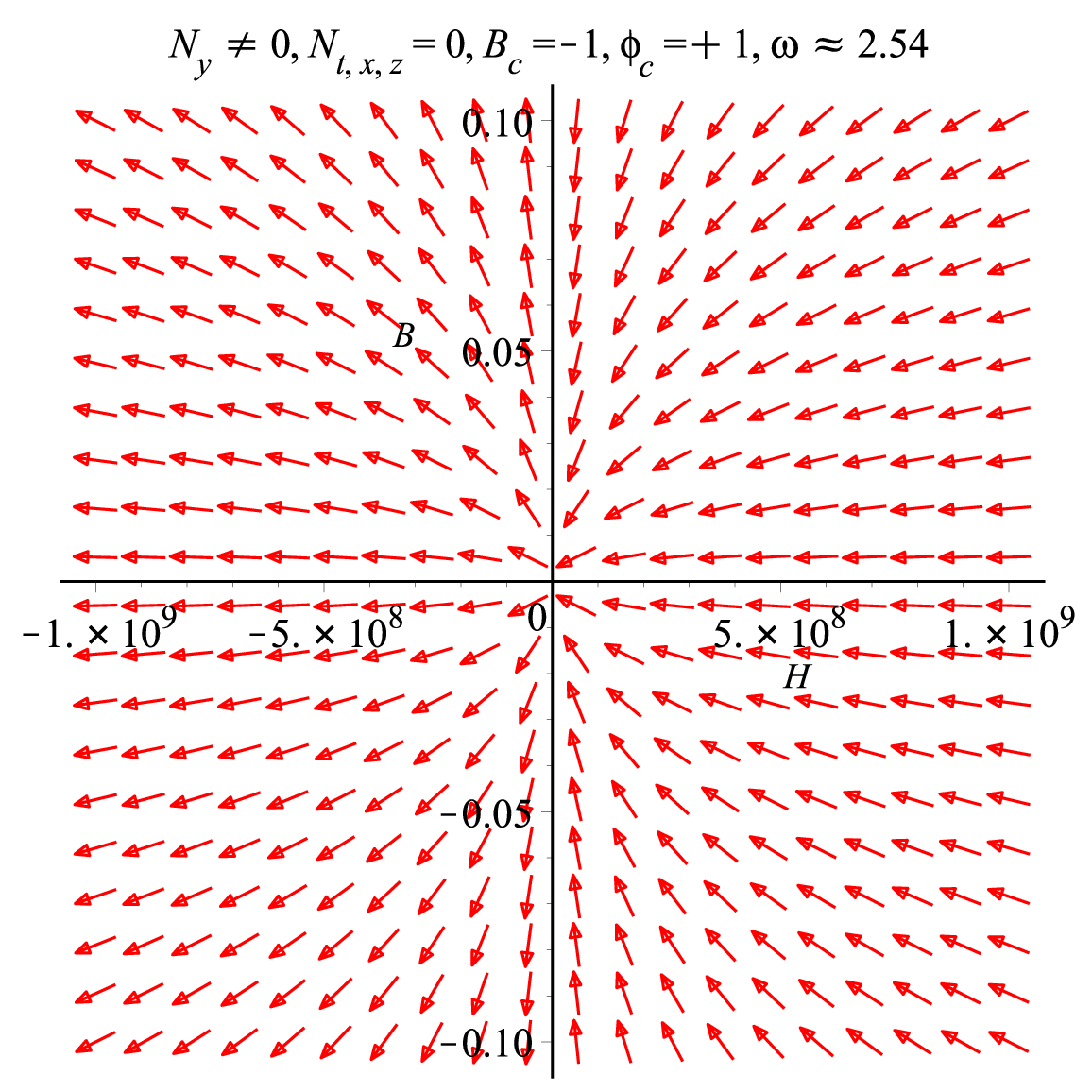}}
 \caption{Arrow diagram for $N_{y}\neq0,~N_{t,x,z}=0$ with quasi stable (saddle) nature} \label{l}
\end{figure}

\end{document}